\documentstyle[12pt]{article}
\newcommand{\bm}[1]{\mbox{\boldmath $#1$}}

\input tcilatex

\begin{document}

\author{V.P.Belavkin \\
%EndAName
University of Nottingham, NG7 2RD, UK \and O. Melsheimer \\
%EndAName
Fachbereich Physik, Renthof 7, Marburg, W-3550, Germany}
\title{A Stochastic Hamiltonian Approach for Quantum Jumps, Spontaneous
Localizations, and Continuous Trajectories. }
\date{20 November 1991. Published in \\
{\it Quantum Semiclass. Opt.} {\bf 8} (1996) 167--187}
\maketitle

\begin{abstract}
We give an explicit stochastic Hamiltonian model of discontinuous unitary
evolution for quantum spontaneous jumps like in a system of atoms in quantum
optics, or in a system of quantum particles that interacts singularly with
''bubbles'' which admit a continual counting observation. This model allows
to watch a quantum trajectory in a photodetector or in a cloud chamber by
spontaneous localisations of the momentums of the scattered photons or
bubbles. Thus, the continuous reduction and spontaneous localization theory
is obtained from a Hamiltonian singular interaction as a result of quantum
filtering, i.e., a sequential time continuous conditioning of an input
quantum process by the output measurement data. We show that in the case of
indistinguishable particles or atoms the a posteriori dynamics is mixing,
giving rise to an irreversible Boltzmann-type reduction equation. The latter
coincides with the nonstochastic Schr\"{o}dinger equation only in the mean
field approximation, whereas the central limit yelds Gaussian mixing
fluctuations described by a quantum state reduction equation of diffusive
type.
\end{abstract}

\begin{description}
\item[Keywords] :\newline
Quantum Jumps, Counting Processes, Continuous Reduction, State Diffusion,
Quantum Trajectories, Spontaneous Localisation.
\end{description}

\section*{Introduction}

\typeout{Introduction} \label{sec:bhaml0} \setcounter{equation}{0}

This paper gives a microscopic theory of quantum jumps, inducing the
spontaneous localizations. It is based on a new type of quantum dynamics,
described by the generalized (singular) stochastic Schr\"{o}dinger equation,
and on the quantum filtering process, amounting to a modification of quantum
evolution along the quantum measurement trajectories. We do not assume the
existance of a unique universal mechanism for continuous reduction and
spontaneous localization during the measurement process. There are many such
mechanisms, as many as the considered kinds of the singular interactions,
amounting to completely unequivalent modifications of quantum evolution
(e.g. quantum jumps and quantum diffusion) along lines, formally similar to
different types of macro-objectification of pointer positions.

The quantum measurement theory based on the ordinary von Neumann reduction
postulate applies neither to instantaneous observations with continuous
spectra nor to continual (continuous in time) measurements. Although such
phenomena can be described in the more general framework of Ludwig's
Davies-Lewies operational approach \nocite%
{bib:bhaml1,bib:bhaml2,bib:bhaml3,bib:XP38,bib:ref33} \cite{bib:bhaml1}--%
\cite{bib:ref33} there is a particular interest in describing quantum
measurements by concrete Hamiltonian models from which the operational
description can be derived by an averaging procedure. Perhaps the first
model of such kind for instantaneous unsharp measurement of particle
localization was given by von Neumann \cite{bib:bhaml6}. He considered the
singular interaction Hamiltonian 
\begin{equation}
h_{x}(t)=x\delta (t)\frac{\hbar }{{\rm i}}\frac{{\rm d}}{{\rm d}\lambda }{,}%
\quad \delta (t)=\left\{ 
\begin{array}{ll}
\infty \,, & t=0 \\ 
0\,, & t\neq 0%
\end{array}%
\right.  \label{eq:bhaml0.1}
\end{equation}%
with Dirac $\delta $-function, producing the translation operator 
\begin{equation}
s_{x}=\exp \biggl\{-\frac{{\rm i}}{\hbar }\,\int_{-\infty }^{\infty }h_{x}(t)%
{\rm d}t\biggr\}=e^{-x{\rm d}/{\rm d}\lambda }  \label{eq:bhaml0.2}
\end{equation}%
at time $t=0$ of the measurement. Here $x$ is the position of the particle
and $\lambda $ is the pointer position, defining, say, the momentum $p$ of a
quantum meter. The particle scattering operator $S=\{s_{x}\}$ applied to the
(generalized) position eigen-kets $|x\rangle $ as $S|x\rangle =|x\rangle
s_{x}$ does not affect the position $x$ of the particle but changes the
meter momentum $p$ to 
\begin{equation}
y=s_{x}^{\dagger }ps_{x}=x+p\,,  \label{eq:bhaml0.10}
\end{equation}%
which implies that the initial wave function $\psi _{0}(x,\lambda )$ of the
system \textquotedblleft particle plus meter\textquotedblright\ is
transformed into 
\begin{equation}
\psi (x,\lambda )=s_{x}\psi _{0}(x,\lambda )=\psi _{0}(x,\lambda -x)\,.
\label{eq:bhaml0.3}
\end{equation}

If in the initial state the particle and the meter were not coherent, $\psi
_0(x,\lambda )=\eta (x)f_0(\lambda )$, and the wave function $f_0$ of the
meter was fixed, then one can obtain the unitary transformation $S:\psi
_0\mapsto s_{\bullet }\psi _0$ via a family $\{F(\lambda )\}$ of reduction
transformations $F(\lambda ):\eta \mapsto f_{\bullet }(\lambda )\eta $ for
the prior particle vector-states $\eta $. Specifically, Eq. (\ref%
{eq:bhaml0.3}) can be defined as 
\begin{equation}  \label{eq:bhaml0.4}
\psi (x,\lambda )=f_0(\lambda -x)\eta (x)\equiv f_x(\lambda )\eta (x)\,.
\end{equation}
The linear nonunitary operators $F(\lambda )=s_{\bullet }f_0(\lambda )$ act
on $|x\rangle $ as the multiplication $F(\lambda )|x\rangle =|x\rangle
f_x(\lambda )$ by $f_x(\lambda )=f_0(\lambda -x)=s_xf_0(\lambda )$ and would
give a sharp localization of any particle wave function $\eta (x)$ at the
point $x=\lambda $ of the pointer position provided that the wave function $%
f_0(\lambda )$ could be initially localized at $\lambda =0$. But there are
no such sharply localized quantum states for the continuous pointer, and the
best that one can do is to take an approximately sharp wave packet $%
f_0(\lambda )$. Say, $f_0$ is the Gaussian packet rescaled to the standard
form 
\begin{equation}  \label{eq:bhaml0.5}
f_0(\lambda )=\exp \biggl\{-\frac \pi 2\,\lambda ^2\biggr\}\,.
\end{equation}
This results in the unsharp localization $[F(\lambda )\eta ](x)=\psi
(x,\lambda )$, 
\begin{equation}  \label{eq:bhaml0.6}
\psi (x,\lambda )=\exp \biggl\{-\frac \pi 2\,(x-\lambda )^2\biggr\}\eta (x)
\end{equation}
of any wave function $\eta (x)$ about the observed value $\lambda $ of the
pointer. The localized wave function $\psi (x,\lambda )$ defines for each
measurement result $\lambda $ the posterior particle vector-state 
\begin{equation}  \label{eq:bhaml0.8}
\eta _\lambda =c_\lambda ^{-1}\psi (\lambda )\,,\quad |c_\lambda |^2=\Vert
\psi (\lambda )\Vert ^2
\end{equation}
up to the normalization 
\begin{equation}  \label{eq:bhaml0.9}
\Vert \psi (\lambda )\Vert ^2=\int |\psi (x,\lambda )|^2{\rm d}x={\rm p}%
(\lambda )
\end{equation}
to the probability density 
\begin{equation}  \label{eq:bhaml0.7}
{\rm p}(\lambda )=\int |f_0(\lambda -x)|^2|\eta (x)|^2{\rm d}x\,.
\end{equation}

For commuting operators $x$ and $y$, this is equivalent to the classical
measurement model $y=x+p$ for unsharp measurement of an unknown signal $x$
via the sharp measurement of the signal plus Gaussian noise $p$ with given
probability density $|f_0(\lambda )|^2$ of the random values $p=\lambda $.

These arguments illustrate how to interpret the reduction model involving
the continuous spectrum of a quantum measurement as a Hamiltonian
interaction model with nondemolition observation for a quantum object using
the measurement of the pointer coordinate of the quantum meter. Since von
Neumann introduced this approach, they were used by numerous other authors 
\cite{bib:bhaml7,bib:bhaml8} for the derivation of a generalized reduction $%
\eta \mapsto F(\lambda )\eta $ that would replace the von Neumann postulate $%
\eta \mapsto E(\lambda )\eta $ given by orthoprojections $\{E(\lambda )\}$
in the case of discrete values $\lambda $.

As extended to nondemolition observations continual in time \nocite%
{bib:ref17,bib:ref18,bib:ref25,bib:grand2,bib:ref19,bib:ref21} \cite%
{bib:ref17}--\cite{bib:grand11}, this idea constitute an essence of the
quantum filtering method for the derivation of nonlinear stochastic wave
equations describing the quantum dynamics under the observation. Since a
particular type of such equations has been taken as a postulate in the
phenomenological theory of permanent reduction, quantum jumps and
spontaneous localization \nocite%
{bib:bhaml16,bib:bhaml17,bib:bhaml18,bib:bel7} \cite{bib:bhaml16}--\cite%
{bib:bel10}, the question arises whether it is possible to obtain this
equation from an appropriate Schr\"odinger equation. Here we shall show how
this can be done by second quantisation of the singular interaction
Hamiltonian considered by von Neumann obtaining a stochastic model of
continual nondemolition observation for the position of a quantum particle
by counting some other quanta.

First we show in the Sec.1 that even the projection postulate can be derived
in the framework of this approach with the suggested Hamiltonian interaction
and a proper nondemolition observation with a discrete spectrum and sharp
initial state of the meter. Then we define a single-kick stochastic wave
equation for the reduced state-vector, corresponding to the unsharp $%
f_0(\lambda )$.

In the Sec.2 we develop the differential treatment of discontinuous unitary
evolution in terms of generalized Schr\"odinger equation corresponding to
the scatterings at given or randomly distributed time instants. Then we show
how it can be reduced to the many-kick stochastic wave equation, describing
spontaneous localization of a quantum particle under the continual
observation of its trajectory in a bubble chamber.

The quantum system of many similar interacting particles in a bubble chamber
is treated in the Sec.3. We prove that the reduced dynamics of the particles
is mixing under the continual observation of positions of the scattered
bubbles. It can not be described by any reduction equation for the posterior
wave function but can be described by an irreversible stochastic equation
for the density matrix of the particles.

Finally we obtain in the Sec.4 the macroscopic and diffusion limits of the
generalized Schr\"odinger and reduction equations under the limit of weak
coupling constant and higher frequency of the interaction with the
apparatus. In the macroscopic limit these two equations coincide with the
ordinary Schr\"odinger equation, corresponding to the mean field
approximation, while in the diffusion approximation, which describes the
fluctuations, they lead to the essentially different kind of stochastic wave
equations \cite{bib:bhaml19, bib:bhaml20}.

These results are presented by the equations (1.10), (2.10), (3.10) and
(4.10).

\section{Hamiltonian Model for a Generalized Reduction}

\typeout{Hamiltonian Model for a Generalized Reduction} \label{sec:bhaml1} %
\setcounter{equation}{0}

Let ${\cal H}$ be a Hilbert space called the state space of a particle, and
let $R$ be a selfadjoint operator in ${\cal H}$ with either integer or
continuous spectrum ${\bf Z}$ or ${\bf R}$. Let $\kappa >0$ be a scaling
parameter. One can regard the scaled operator $\kappa R$ as the position $x$
of the particle in ${\bf R}$ or in the lattice $\kappa {\bf Z}$ if it is
quantised so that $\kappa R$ is given in the $x$-representation, the
operator acts as the multiplication $\kappa R|x\rangle =|x\rangle \lambda
(x) $ by $\lambda (x)=\kappa \lfloor x/\kappa \rfloor $, where $\lfloor
x\rfloor \in {\bf Z}$ denotes the integer part of $x$.

A quantum meter with continuous ($\Lambda ={\bf R}$) or discrete ($\Lambda
=\varepsilon {\bf Z}$) pointer scale is described by the Hilbert space $%
L^{2}(\Lambda )$ of complex-valued functions $f:\Lambda \rightarrow {\bf C}$
square integrable in the sense that $\Vert f\Vert ^{2}=\int |f(y)|^{2}{\rm d}%
y<\infty $. The last integral coicides with $\sum_{y\in \Lambda
}|f(y)|^{2}\varepsilon $ if $f(y)=\sum_{l\in {\bf Z}}\delta
_{l}^{\varepsilon }(y)f(\varepsilon l)\varepsilon $ is the isometric
interpolation of the discrete wave-function $f(\varepsilon l)$, $l\in {\bf Z}
$, given by 
\[
\delta _{l}^{\varepsilon }(y)=\int_{0}^{1/\varepsilon }e^{2\pi {\rm i}%
k(\varepsilon l-y)}{\rm d}k\,,\quad y\in {\bf R}\,. 
\]

Consider a moving particle with Hamiltonian $H$ in ${\cal H}$. Its singular
evolution corresponding to the position measurement at time $t=0$ is
described in the product space ${\cal H}_{1}={\cal H}\otimes L^{2}(\Lambda )$
by time-dependent Hamiltonian 
\begin{equation}
H_{1}(t)=H\otimes {\bf 1}-\kappa R\otimes \delta (t)Q\,,  \label{eq:bhaml1.1}
\end{equation}%
in the interaction picture corresponding to the free translation evolution
of the poiter coordinate $p\in \Lambda $. Here ${\bf 1}$ is the identity
operator in $L^{2}(\Lambda )$, defining the particle Hamiltonian $%
H_{0}=H\otimes {\bf 1}$ in ${\cal H}_{1}$, and $Q$ is treated as a
coordinate operator of the meter, $Q={\rm i}\hbar \partial /\partial y$,
having the bound spectrum $2\pi \hbar \lbrack 0,1/\varepsilon )$ for $%
\Lambda =\varepsilon {\bf Z}$. This generates the shift operator (\ref%
{eq:bhaml0.2}) in $L^{2}(\Lambda )$ with $x\in \Lambda $ and the scattering
operator 
\begin{equation}
S_{t}=\exp \biggl(\frac{{\rm i}}{\hbar }\kappa R\otimes Q_{t}\biggr)=\left\{ 
\begin{array}{ll}
S\,, & t>0 \\ 
I\,, & t\leq 0%
\end{array}%
\right.  \label{eq:bhaml1.2}
\end{equation}%
in ${\cal H}_{1}$, where $Q_{t}=1_{t}Q$, $1_{t}=1$ if $t>0$, and $1_{t}=0$
otherwise.

The singular time dependence of the Hamiltonian (\ref{eq:bhaml1.1}) makes it
impossible to define the Schr\"{o}dinger equation ${\rm i}\hbar {\rm d}\psi /%
{\rm d}t=H_{1}(t)\psi (t)$ at $t=0$ in the usual sense. But one can define a
discontinuous unitary evolution $U_{1}(t):{\cal H}_{1}\rightarrow {\cal H}%
_{1}$ for some $t_{0}<0$ as the single-jump unitary process 
\[
U_{1}(t)=\exp \biggl(-\frac{{\rm i}}{\hbar }\int_{t_{0}}^{t}H_{1}(s){\rm d}s%
\biggr)=e^{{\rm i}H_{0}(t_{0}-t)/\hbar }S_{t} 
\]%
provided that $[R,H]=0$. If $R$ and $H$ do not commute, the usual method
which a physicists would follow at facing this difficulty is to consider the
free evolution $\psi _{-}(t)$ for $t<0$ and $\psi _{+}(t)$ for $t>0$,
supplemented by the boundary condition $\psi _{+}(0)=S\psi _{-}(0)$.
Although this method is hardly applicable in the case of spontaneous
interaction at a random instant $t_{1}>t_{0}$, let us show that the
discontinuous unitary evolution corresponding to the fixed $t_{1}=0$ can be
defined as the fundamental solution $\psi (t)=U_{1}(t)\psi _{0}$ of a
generalized Schr\"{o}dinger equation. Such regularized equation can be
written in terms of the forward differentials 
\[
{\rm d}\psi (t)=\psi (t+{\rm d}t)-\psi (t),\quad {\rm d}1_{t}=1_{t+{\rm d}%
t}-1_{t}\,, 
\]%
as the Ito integral of 
\begin{equation}
{\rm d}\psi (t)+\frac{{\rm i}}{\hbar }H_{0}\psi (t){\rm d}t=(S-I)\psi (t)%
{\rm d}1_{t}\,,\quad \psi (t_{0})=\psi _{0}\,.  \label{eq:bhaml1.3}
\end{equation}

For every initial $\psi _{0}\in {\cal H}$ the equation (\ref{eq:bhaml1.3})
has a unique solution at $t_{0}\leq 0$; this solution is given by the
unitary operator $U_{1}(t)=U_{0}(t-t_{0})S_{t}(-t_{0})$, where $%
U_{0}(t)=\exp \{-{\rm i}H_{0}t/\hbar \}$, $S_{t}(r)=U_{0}^{\dagger
}(r)S_{t}U_{0}(r)$.

To prove this let us rewrite the generalized Schr\"{o}dinger equation in the
integral form 
\begin{equation}
\psi (t)=e^{-{\rm i}H_{0}t/\hbar }\biggl(e^{{\rm i}H_{0}t_{0}/\hbar }\psi
_{0}+\int_{t_{0}}^{t}e^{{\rm i}H_{0}r/\hbar }(S-I)\psi (r){\rm d}1_{r}\biggr)
\label{eq:bhaml1.4}
\end{equation}%
(the equivalence of (\ref{eq:bhaml1.4}) to (\ref{eq:bhaml1.3}) can be shown
by straightforward differentiation). We can write $\psi
(0)=U_{0}(-t_{0})\psi _{0}$ for the solution (\ref{eq:bhaml1.4}) when $t=0$
since $\int_{t_{0}}^{0}\varphi (r){\rm d}1_{r}=0$ if $t_{0}\leq 0$; hence $%
\psi (t)$ can be rewritten as 
\[
\psi (t)=U_{0}(t)(\psi (0)+(S-I)1_{t}\psi (0))=U_{0}(t)S_{t}\psi (0)\,. 
\]%
It follows from $\int_{t_{0}}^{t}\varphi (r){\rm d}1_{r}=1_{t}\varphi (0)$
for any function $\varphi $ and any $t_{0}<0$, and $(S-I)1_{t}=S_{t}-I$ by
the definition of the scattering operator (\ref{eq:bhaml1.2}). Thus, $%
U_{1}(t)$ is the unitary operator 
\[
U_{0}(t)S_{t}U_{0}(-t_{0})=U_{0}(t-t_{0})S_{t}(-t_{0}). 
\]%
This gives the solution to equation (\ref{eq:bhaml1.3}) for $t_{0}=0$ as
well, since $\psi (t)=U_{0}(t)S_{t}\psi (0)$ is equal to $\psi _{0}$ for $%
t=0 $.

The rescaled pointer operator $P=\kappa ^{-1}I\otimes p$, switched on at the
instant $t=0$ of the scattering, is described in the space ${\cal H}\otimes
L^{2}(\Lambda )$ by the operator $P_{t}=1_{t}P$, having in the Heisenberg
picture $Y_{t}=U_{1}^{\dagger }(t)P_{t}U_{1}(t)$ the constant form $%
Y_{t}=Y(-t_{0})$ for all $t>0$. Here $Y(r)=S^{\dagger }(r)PS(r)$ taken at $%
r=-t_{0}$, $t_{0}\leq 0$ is the shifted operator 
\begin{equation}
Y(r)=R(r)\otimes {\bf 1}+\frac{1}{\kappa }\,I\otimes p\,,
\label{eq:bhaml1.5}
\end{equation}%
where $R(r)=U^{\dagger }(r)RU(r)$ and $S(r)=\exp (\kappa R(r)\otimes {\rm i}%
Q/\hbar )$. Obviously the observables $\{Y_{t}\}$ are self-nondemolition in
the sense of their joint measurability. It follows from the trivial
commutativity condition 
\begin{equation}
\lbrack Y_{s},Y_{t}]=0\,,\quad \forall \,s,t\,.  \label{eq:bhaml1.6}
\end{equation}%
And they are nondemolition with respect to an arbitrary particle operator $%
X_{t}(r)=U_{1}(t)^{\dagger }(X\otimes {\bf 1})U_{1}(t)$ in the Heisenberg
picture at $t_{0}=-r$ in the sense of their predictability \cite%
{bib:ref17,bib:ref19} 
\begin{equation}
\lbrack X_{s},Y_{t}]=0\,,\quad \forall \,s\geq t,  \label{eq:bhaml1.7}
\end{equation}%
indeed, $[X_{s},Y_{s}]=0$ and $Y_{s}=Y_{t}$ for $s,\,t>0$ since 
\[
U_{1}^{\dagger }(s)P_{t}U_{1}(s)=S^{\dagger }(-t_{0})P_{t}S(-t_{0})\,,\quad
\forall \,s\geq t\,. 
\]

Let us fix a state vector $f_{0}\in L^{2}(\Lambda )$, $\Vert f_{0}\Vert =1$,
given by a wave function $f_{0}(y)$ on $\Lambda $ localized at $y=0$. Let $%
|y)$ denote the (generalized) eigen function in the spectral representation $%
p=\int y|y)(y|{\rm d}\lambda $ of the pointer coordinate $(pf)(y)=yf(y)$,
where ${\rm d}\lambda ={\rm d}y$ if $y\in {\bf R}$ and ${\rm d}\lambda
=\varepsilon $ if $y\in \varepsilon {\bf Z}$. This yields the localizing
transformations $F_{t}(y)=(y|S_{t}f_{0}$ in the form 
\begin{equation}
F_{t}(y)=f_{0}(yI-\kappa R_{t})=\left\{ 
\begin{array}{ll}
F(y)\,, & t>0 \\ 
f_{0}(y)I\,, & t\leq 0%
\end{array}%
\right.  \label{eq:bhaml1.8}
\end{equation}%
where $F(y)=f_{0}(yI-\kappa R)$. The reduction transformations $\eta \mapsto
\psi (t,y)$, defined on the particle space ${\cal H}$ by the formula 
\[
\psi (t,y)=U(t-t_{0})F_{t}(-t_{0},y)\eta \,,\quad y\in \Lambda , 
\]%
with $U(t)=\exp (-{\rm i}Ht/\hbar )$ and $F_{t}(r)=U^{\dagger }(r)F_{t}U(r)$
reproduce the unitary evolution $U_{1}(t)$ in the $y$-representation on $%
\eta \otimes f_{0}\in {\cal H}_{1}$ similarly to Eq. (\ref{eq:bhaml0.3}) and
(\ref{eq:bhaml0.4}), namely, 
\[
U(t-t_{0})F_{t}(-t_{0},y)\eta =U(t)(y|S_{t}f_{0}U(-t_{0})\eta
=(y|U_{1}(t)(\eta \otimes f_{0})\,. 
\]

The operator $F(r,y)=f_0(yI-\kappa R(r))$ at $r=-t_0$ with a given initial
vector-state $\eta $ of the particle before the scattering ($t_0\leq 0$)
define the probability measure 
\[
\mu (\Delta )=\int_\Delta \Vert F(r,y)\eta \Vert ^2{\rm d}\lambda =\langle
\eta ,\Pi _t(\Delta )\eta \rangle \,,\quad \Delta \subseteq \Lambda \,, 
\]
for the statistics of the nondemolition measurement of $\kappa R$ via the
observation of the pointer position $y\in \Delta $ after the scattering. It
is given by a positive operator-valued measure $\Pi _t(\Delta )=\int_\Delta
|F_t(y,-t_0)|^2{\rm d}\lambda $ at $t>0$. (Before the scattering it defines
the initial probability measure 
\[
\mu _0(\Delta )=\langle \eta ,\Pi _t(\Delta )\eta \rangle =\int_\Delta
|f_0(y)|^2{\rm d}\lambda \,,\quad t\leq 0\ 
\]
which is independent of $\eta $.)

In the case of discrete scale the eigen functions $|y)$ are normalisable: $%
(y|y)=1/\varepsilon $, and one can take a sharply localized $f_0=\varepsilon
^{d/2}|0)$, given by the eigen function $f_0(y)=e(y)/\varepsilon ^{d/2}$,
where $e(y)=1$ if $y=0$, $e(y)=0$ if $y\neq 0$. By renormalising the
operator $F(y)$ as $E(y)=\varepsilon ^{d/2}F(y)$ with step $\varepsilon
=\kappa $ of the position quantisation, one obtain the orthogonal
projections 
\[
e(\kappa R-yI)=\int_{x:\lfloor x/\kappa \rfloor =y/\kappa }|x\rangle \langle
x|{\rm d}x=E(y)\,,\quad y\in \kappa {\bf Z} 
\]
of $\kappa R=\int \lambda (x)|x\rangle \langle x|{\rm d}x=\sum_{y\in \kappa 
{\bf Z}}yE(y)$ corresponding to eigen values $\lambda (x)=\kappa \lfloor
x/\kappa \rfloor $. Thus, $\Pi _t(\Delta )$, $\Delta \subseteq \Lambda $ is
the spectral measure $\sum_{y\in \Delta }E(t_0,y)$ of the quantised position 
$\kappa R(-t_0)$ of the particle for a $t>0$ given by the eigen
orthoprojections $E(r,y)=U^{\dagger }(r)E(y)U(r)$ of the operator $R(r)$,
corresponding to the rescaled pointer integer values $y/\kappa $. Thus, the
projection reduction postulate has been deduced from the Hamiltonian
interaction (\ref{eq:bhaml1.1}) and the nondemolition measurement for the
sharply localized initial state $f_0$. But there is no continuous limit as $%
\kappa \rightarrow 0$ of such sharp reduction with nontrivial $e\neq 0$,
since $\Vert e\Vert ^2=\int |e(y)|^2{\rm d}\lambda =\kappa \rightarrow 0$
and the sharp function $e(y)=\delta _0^\kappa (y)\kappa $ on $\Lambda
=\kappa {\bf Z}$ disappears as an element of the state space $L^2({\bf R})$
of the continuous meter.

In the continuous case, one can take only an approximately sharp $f_0\in L^2(%
{\bf R})$, say $f_0(y)=\delta _0^\varepsilon (y)\varepsilon $, $y\in {\bf R}$%
, and renormalise the operator (\ref{eq:bhaml1.8}) as $G_t(y)=f_0(yI-\kappa
R_t)/f_0(y)$ if $f_0(y)\neq 0$, as in the Gaussian case (\ref{eq:bhaml0.5}).
They define $G_t(y)$ as the identity operator for $t\leq 0$ or $r>0$,
whereas for $t>0$ and $r\leq 0$ the operator 
\begin{equation}
G(y)=\langle y|Sf_0=(y|G\,,\quad G=f_0^{-1}Sf_0\,,  \label{eq:bhaml1.9}
\end{equation}
say, of the Gaussian form 
\[
G(y)=\exp \{\pi \kappa R(yI-\frac 12\,\kappa R)\} 
\]
given by the generalized eigen functions $|y\rangle =|y)/f_0(y)$ for the
spectral representation $p=\int y|y\rangle \langle y|{\rm d}\mu _0$ with
respect to the initial probability measure ${\rm d}\mu _0=|f_0(y)|^2{\rm d}%
\lambda $ with density $|f_0(y)|^2=\exp \{-\pi y^2\}$ in the Gaussian case.

The corresponding propagators 
\[
T(t,y)=U(t-t_0)G_t(-t_0,y),\quad y\in \Lambda ,\;t_0\leq 0\,, 
\]
where $G_t(r)=U^{\dagger }(r)G_tU(r)$, define the operator--valued measure $%
\Pi _t(\Delta )$ as 
\[
\Pi _t(\Delta )=\int_\Delta T^{\dagger }(t,y)T(t,y){\rm d}\mu _0=\int_\Delta
|G_t(-t_0,y)|^2{\rm d}\mu _0 
\]
so that the output probability measure $\mu $ is absolutely continuous with
respect to $\mu _0$ for any $\eta \in {\cal H}$: $\mu (\Delta )=\int_\Delta
\Vert G(-t_0,y)\eta \Vert ^2{\rm d}\mu _0$. Hence, the reduced state vector $%
\chi (t,y)=T(t,y)\eta $ is normalized to $1$ as a stochastic vector process $%
\chi (t):y\mapsto \chi (t,y)\in {\cal H}$ in the mean square sense with
respect to the input probability measure $\mu _0$, 
\[
\Vert \chi (t)\Vert _0^2=\int \langle \chi (t,y),\chi (t,y)\rangle {\rm d}%
\mu _0=\langle \eta ,\eta \rangle =1\,. 
\]
This model of nondemolition observation with continuous data $y\in {\bf R}$
also applies to unsharp measurement of operator $R$ with discrete spectrum.
In contrast to sharp measurement, unsharp measurement is not sensitive to
the continuous spectrum limit as $\kappa \rightarrow 0$ of $R=\lfloor
x/\kappa \rfloor $, corresponding to the replacement of $\kappa R$ by $x\in 
{\bf R}$.

For any initial $t_{0}\leq 0$ the stochastic vector process $\chi
(t)=T(t)\eta $ satisfies the single-kick equation 
\begin{equation}
{\rm d}\chi (t)+\frac{{\rm i}}{\hbar }H\chi (t){\rm d}t={\rm d}%
1_{t}[G-I]\chi (t)\,,\quad \chi (t_{0})=\eta  \label{eq:bhaml1.10}
\end{equation}%
generated by the random differential ${\rm d}1_{t}[G-I](y)=(G(y)-I){\rm d}%
1_{t}$ on $y\in \Lambda $ with respect to the initial probability measure $%
\mu _{0}$. This simplest reduction equation is written in terms of the
forward differentials ${\rm d}\chi (t,y)=\chi (t+{\rm d}t,y)-\chi (t,y)$,
that is, is understood in the sense of Ito.

Indeed, by representing $G_{t}$ as $G_{t}(y)=(G(y)-I)1_{t}+I=I+1_{t}[G-I](y)$%
, one can describe $\chi (t)=U(t-t_{0})G_{t}(-t_{0})\eta $ by the integral
equation 
\begin{eqnarray*}
\lefteqn{\chi (t)=U(t-t_{0})\eta +U(t)1_{t}[G-I]U(-t_{0})\eta =} \\
&=&e^{-{\rm i}Ht/\hbar }\biggl(e^{{\rm i}Ht_{0}/\hbar }\eta
+\int_{t_{0}}^{t}e^{{\rm i}Hr/\hbar }{\rm d}1_{r}[G-I]\chi (r)\biggr)\,.
\end{eqnarray*}%
But this equation is equivalent to the differential equation (\ref%
{eq:bhaml1.10}), a fact that can be proved by straightforward
differentiation taking into account the Ito multiplication table 
\[
({\rm d}t)^{2}=0\,,\quad {\rm d}t{\rm d}1_{t}=0={\rm d}1_{t}{\rm d}t\,,\quad
({\rm d}1_{t})^{2}={\rm d}1_{t}\,. 
\]

Similarly, one can the simplest nonlinear stochastic equation for the
normalized reduced state vector $\chi _{y}(t)=\chi (t,y)/\Vert \chi
(t,y)\Vert $: 
\[
{\rm d}\chi _{y}(t)+\frac{{\rm i}}{\hbar }H\chi _{y}(t){\rm d}%
t=(G_{y}(t)-I)\chi _{y}(t){\rm d}1_{t}\,,\quad \chi _{y}(0)=\eta 
\]%
where $G_{y}(t)=G(y)/\Vert G(y)\chi _{y}(t)\Vert $, $\eta \in {\cal H}$.
This equation is an equivalent differential form the nonlinear integral
stochastic equation 
\begin{eqnarray*}
\chi _{y}(t) &=&e^{-{\rm i}Ht/\hbar }\biggl(e^{{\rm i}Ht_{0}/\hbar }\eta
+\int_{t_{0}}^{t}e^{{\rm i}Hr/\hbar }(G_{y}(r)-I)\chi _{y}(r){\rm d}1_{r}%
\biggr)= \\
&=&U(t-t_{0})\eta +U(t)(G_{y}-I)1_{t}\chi _{y}(0)=U(t)G_{t}(y)\chi
_{y}(0)/\Vert G_{t}(y)\chi _{y}(0)\Vert \,.
\end{eqnarray*}%
This yields $\chi _{y}(t)=T(t,y)\eta /\Vert T(t,y)\eta \Vert $ for a $%
t_{0}\leq 0$ because of $\Vert G_{t}(y)\eta \Vert =\Vert T(t,y)\eta \Vert $.

Note that the random state vector $\chi _y(t)$ is obtained by conditioning
with respect to the output (rather than input) probability measure ${\rm d}%
\mu =\Vert \chi (t,y)\Vert ^2{\rm d}\mu _0$.

\section{Spontaneous Localization of a Single Particle}

\typeout{Spontaneous Localization of a Single Particle} \label{sec:bhaml2} %
\setcounter{equation}{0}

Let us consider a spontaneous process of scattering interactions (\ref%
{eq:bhaml1.1}) of a quantum particle (or an atom) at random time instants $%
t_n>0$, $t_1<t_2<\ldots $, with a renewable meter in an apparatus of the
cloud chamber (photodetector) type with bubbles (photons) serving as the
meter. We consider the increasing sequences $(t_1,t_2,\ldots )$ as countable
subsets $\tau \subset {\bf R}_{+}$ such that $\tau _t=\tau \cap [0,t)$ is
finite for any $t\geq 0$ in accordance with the finiteness of the number of
scattered bubbles on the finite observation interval $[0,t)$. The set of all
such infinite $\tau $ will be denoted by $\Gamma _\infty $, and $\Gamma $ is
the inductive limit $\cup \Gamma _t$ as $t\rightarrow \infty $ of $\Gamma
_t=\{\tau _t|\tau \in \Gamma _\infty \}$, which is the disjoint union $%
\Gamma _t=\sum_{n=0}^\infty \Gamma _t(n)$ of $n$-simplice $\Gamma
_t(n)=\{t_1<\ldots <t_n\}\subset [0,t)^n$.

The measurement apparatus is assumed to be a quantum system of infinitely
many bubbles each of which is identical to the single meter described in the
previous section. The pointer coordinate is attached to the momentum $p_n$
of a bubble labeled by the scattering number $n\in {\bf N}$, such that it
shows the momentum $\lambda _n\in \Lambda $ of the scattered bubble at each
time $t_n\in \tau $.

The corresponding Hamiltonian of the moving particle is given by the series 
\begin{equation}
H(t,\tau )=H_{0}-\kappa R\otimes \sum_{n=1}^{\infty }\delta (t-t_{n})Q(n)
\label{eq:bhaml2.1}
\end{equation}%
having at most two nonzero terms if $t\in \tau $. Here $H_{0}=H\otimes {\bf 1%
}^{\otimes \infty }$ is the Hamiltonian describing the time evolution on the
intervals between the scatterings $t\in \tau $ and $Q(n)$ is the coordinate
of the $n$th scattered bubble, given as the operator $Q(n)={\rm i}\hbar {\rm %
d}/{\rm d}\lambda _{n}$.

The generalized Schr\"{o}dinger equation corresponding to the Hamiltonian (%
\ref{eq:bhaml1.1}) can be written for fixed $\tau \in \Gamma _{\infty }$ by
analogy with the single-kick case 
\begin{equation}
{\rm d}\psi (t)+\frac{{\rm i}}{\hbar }\,H_{0}\psi (t){\rm d}%
t=(S(n_{t})-I)\psi (t){\rm d}n_{t},\quad \psi (0,\tau )=\psi _{0}\,.
\label{eq:bhaml2.2}
\end{equation}%
Here $S(n)=\exp \{\frac{{\rm i}}{\hbar }\kappa R\otimes Q(n)\}$ and $%
n_{t}(t)=|\tau _{t}|$ is the numerical process that gives the cardinality $%
|\tau _{t}|=\sum_{r\in \tau }1_{t-r}$ of the localized subset $\tau
_{t}=\{t_{n}<t\}$, so that ${\rm d}n_{t}(\tau )$ is equal to $1$ for $t\in
\tau $, and zero otherwise.

The solution to this equation is uniquely determined for every $\tau \in
\Gamma _{\infty }$ by the initial state $\psi _{0}$ of the system. Namely, 
\[
\psi (t,\tau )=U(t,\tau )\psi _{0}\,, 
\]%
where $U(t,\tau )=U_{0}(t)V_{t}^{\dagger }(\tau )$, $V_{t}^{\dagger }(\tau )$
is the chronological product $\prod_{r\in \tau }^{\leftarrow
}S_{t}(r)=S(t_{n_{t}})\ldots S(t_{1})$, and 
\begin{equation}
V_{t}(\tau )=S_{t}^{\dagger }(t_{1})S_{t}^{\dagger }(t_{2})\ldots =\biggl(%
\prod_{r\in \tau }^{\leftarrow }S_{t}(r)\biggr)^{\dagger }\,.
\label{eq:bhaml2.3}
\end{equation}%
Here $S_{t}(t_{n})=U_{0}^{\dagger }(t_{n})S_{t}(n)U_{0}(t_{n})$ for $t_{n}<t$%
, where $S(n)=\exp \{-\frac{{\rm i}}{\hbar }\,\kappa R\otimes Q(n)\}$, and $%
S_{t}(t_{n})=I$ if $t_{n}\geq t$ so that the infinite product (\ref%
{eq:bhaml2.3}) contains only a finite number $n_{t}=\sum_{r\in \tau }1_{t-r}$
of factors different from the identity operator $I$.

Recall that the differential equation (\ref{eq:bhaml2.2}) is equivalent to
the integral equation given by the recurrence relation 
\begin{equation}
\psi (t,\tau )=e^{-{\rm i}H_{0}t/\hbar }\biggl(\psi _{0}+\sum_{r\in \tau
}^{r<t}e^{{\rm i}H_{0}r/\hbar }(S(n_{r})-I)\psi (r,\tau )\biggr)
\label{eq2.4}
\end{equation}%
for every $\tau \in \Gamma _{\infty }$. Hence, $\psi (t,\tau
)=U_{0}(t)V_{t}^{\dagger }(\tau )\psi _{0}$, where $U_{0}(t)=e^{-{\rm i}%
H_{0}t/\hbar }$ and $V_{t}(\tau )$ is a solution to the operator equation 
\[
V_{t}(\tau )=I+\sum_{r\in \tau }^{r<t}V_{r}(t)(S_{t}^{\dagger }(r,\tau
)-I)\,,\quad V_{0}(\tau )=I\,, 
\]%
where $S^{\dagger }(t,\tau )=U_{0}(t)^{\dagger }S(n_{t}(\tau ))^{\dagger
}U_{0}(t)$. But this equation has a unique solution (\ref{eq:bhaml2.2}),
which can be written as the binomial sum 
\[
\lbrack L_{t}(t_{1})+I][L_{t}(t_{2})+I]\ldots =\sum_{\sigma \subseteq \tau
_{t}}L(s_{1},\tau )\ldots L(s_{n},\tau ) 
\]%
in terms of $\sigma =\{s_{1},\ldots ,s_{n}\}$, $s_{1}<\ldots <s_{n}$, $n\leq
n_{t}$, $L_{t}(r)=S_{t}^{\dagger }(r)-I$ ($=0$ if $r\geq t$) and $L(r,\tau
)=S^{\dagger }(r,\tau )-I$. Indeed, this sum contains $I$ as the null
product corresponding to $\sigma =\emptyset $, and the sum of the other
terms is equal to 
\begin{eqnarray*}
V_{t}(\tau )-I &=&\sum_{r\in \tau }^{r<t}\sum_{\sigma \subseteq \tau
_{r}}L(s_{1},\tau )\ldots L(s_{m},\tau )L(r,\tau )= \\
&=&\sum_{r\in \tau }^{r<t}V_{r}(\tau )L(r,\tau )=\sum_{r\in \tau
}^{r<t}V_{r}(\tau )(S^{\dagger }(r,\tau )-I)\,,
\end{eqnarray*}%
where $m\leq n_{t}-1$.

Note that the generalized differential equation (\ref{eq:bhaml2.2})
depending on $\tau \in \Gamma _\infty $ via $n_t=n_t(\tau )$ is not
necessarily stochastic as long as we have not fixed a probability
distribution for the instants $\tau =(t_1,t_2,\ldots )$ of the singular
interactions. In order to obtain a continuous (at least in the mean)
dynamics for such a quantum jump process it is necessary to assume the
interactions are spontaneous with a continuous probability distribution of
random instants $\tau $. One can assume that the number process $n_t(\tau )$
is stochastic, given by the Poisson law $\pi _0({\rm d}\tau )$ on $\Gamma
_\infty $ presented as the projective limit as $t\rightarrow \infty $ of the
probability measures 
\begin{equation}  \label{eq:bhaml2.5}
\pi _0({\rm d}\tau _t)=e^{-\nu t}\nu ^{|\tau _t|}{\rm d}\tau _t\,,\quad \nu
>0\,.
\end{equation}
Here $\tau _t=\tau $ is a finite time--ordered sequence $\tau
(n)=(t_1,\ldots ,t_n)\in \Gamma _t$ with $n=n_t$, ${\rm d}\tau
_t=\prod_{k=1}^{n_t}{\rm d}t_k$ is the measure on $\Gamma _t$ given by the
sum of product measures ${\rm d}t_1,\ldots ,{\rm d}t_n={\rm d}\tau (n)$ on
the simplices $\Gamma _t(n)$, ${\rm d}\tau (0)=1$ on $\Gamma
_t(0)=\{\emptyset \}$ such that 
\[
\int_{\Gamma _t}\nu ^{|\tau |}{\rm d}\tau \negthinspace
:\,=\sum_{n=0}^\infty \nu ^n\int \ldots \int_{0\leq t_1<\ldots <t_n<t}{\rm d}%
t_1\ldots {\rm d}t_n=e^{\nu t}\,. 
\]
Note that any other numerical process can be described by a positive density
function $f(\tau )$ with respect to the Poissonian measure, that is, has the
form $f(\tau )\pi ({\rm d}\tau )$.

Let us fix an initial state $\varphi _0=f_0^\infty $ of the bubbles as the
infinite product $f_0^\infty =\otimes _{k=1}^\infty f_k$ of the identical
state vectors $f_k=f_0$ of the bubbles given by a normalized element $f_0\in
L^2(\Lambda )$. This defines the solutions $\psi (t,\tau )$ of the
stochastic equation (\ref{eq:bhaml2.2}) with the initial data $\psi _0=\eta
\otimes \varphi _0$ given by state vectors $\eta \in {\cal H}$ in the
particle space ${\cal H}$ as the state vectors in the product space ${\cal H}%
_\infty ={\cal H}\otimes {\cal E}$, where ${\cal E}$ is the Hilbert space
generated by the infinite-product functions $\varphi (\upsilon
)=\prod_{k=1}^\infty f_k(\lambda _k)$ with the equal elements $f_k=f_0$ for
almost all $k$. We suppose, as in Sec.\ref{sec:bhaml1}, that $\varphi
_0(\upsilon )\neq 0$ for almost all $\upsilon =(\lambda _1,\lambda _2,\ldots
)$ such that one can identify the space ${\cal E}$ with the space $%
L_0^2(\Lambda ^\infty )$ of all square-integrable functions $f=\varphi
/\varphi _0$ with respect to the product measure $\mu _0^\infty ({\rm d}%
\mbox{\boldmath $\lambda$})=\mu _0({\rm d}\lambda _1)\cdot \mu _0({\rm d}%
\lambda _2)\cdots $ on the space $\Lambda ^\infty =\Lambda \times \Lambda
\times \ldots $ of the sequences $\mbox{\boldmath $\lambda$}=(\lambda
_1,\lambda _2,\ldots )$: $\Vert f\Vert _0^2=\int |f(%
\mbox{\boldmath
$\lambda$})|^2\mu _0^\infty ({\rm d}\mbox{\boldmath $\lambda$})=\Vert
\varphi \Vert ^2$. The generalized product-vectors $|%
\mbox{\boldmath
$\lambda$}\rangle =|\lambda _1\rangle \otimes |\lambda _2\rangle \otimes
\ldots $ of this space are defined by tensor-product of the $\delta $%
-functions $\langle \lambda ^{\prime }|\lambda \rangle $, respectively to $%
\mu _0$ such that $\int |\mbox{\boldmath $\lambda$}\rangle \langle %
\mbox{\boldmath $\lambda$}|\mu _0^\infty ({\rm d}\mbox{\boldmath $\lambda$})=%
{\bf 1}^{\otimes \infty }$.

Consider the sequence $(p_1,p_2,\ldots )$ of momentums $p_n$, of the
scattered bubbles at the time instants $\{t_1,t_2,\ldots \}$. The commuting
operators $p_n$, $n\in {\bf N}$, described in ${\cal E}$ by the
multiplications $p_n|\mbox{\boldmath $\lambda$}\rangle =|%
\mbox{\boldmath
$\lambda$}\rangle \lambda _n$, are assumed to be measured at the random time
instants $t_n$, $n\in {\bf N}$. The point trajectories of such measurements
are given by the sequences $\mbox{\boldmath $$}=(y_1,y_2,\ldots )$ of pairs $%
y_n=(t_n,\lambda _n)$ with $t_1<t_2,\ldots $ and $\lambda _n\in \Lambda $,
identified with countable subsets $\{y_1,y_2,\ldots \}\subset {\bf R}%
_{+}\times \Lambda $. As elements $\upsilon =(\tau ,%
\mbox{\boldmath
$\lambda$})$ of the Cartesian product $\Upsilon _\infty =\Gamma _\infty
\times \Lambda ^\infty $, they have the probability distribution ${\rm P}_0(%
{\rm d}\upsilon )=\pi _0({\rm d}\tau )\mu _0^\infty ({\rm d}%
\mbox{\boldmath
$\lambda$})$, where $\Lambda ^\infty $ is the space of all sequences $%
\mbox{\boldmath $\lambda$}\in \Lambda ^\infty $, $\lambda _n\in \Lambda $,
equipped with the probability product-measure $\mu _0^\infty ({\rm d}%
\mbox{\boldmath $\lambda$})$.

The measurement data of the observable process up to a given time instant $%
t>0$ is described by a finite sequence $\upsilon _t=(y_1,\ldots ,y_n)$ with $%
n=n_t(\upsilon )$ given by the numerical process $n_t(\tau )$ for the
component $\tau $ of $\upsilon $.

Let us introduce the counting distribution $n_t(\Delta )=|\upsilon _t\cap (%
{\bf R}_{+}\times \Delta )|$ as the number $n_t(\Delta ,\upsilon )$ of
scatterings in the time-space region $[0,t)\times \Delta $ and define the
counting integral $\int_0^\infty \int_\Lambda L(r,\lambda ){\rm d}n_r({\rm d}%
\lambda )$ over $y\in {\bf R}_{+}\times \Lambda $ as the series 
\begin{equation}  \label{eq:bhaml2.6}
n[L](\upsilon )=\sum_{y\in \upsilon }L(y)\,,\quad \forall \,\upsilon \in
\Upsilon _\infty \,.
\end{equation}
Having fixed an integer-valued distribution $n_t(\Delta )\in \{0,1,\ldots \}$
as a function of $t\geq 0$ and of measurable sets $\Delta \subseteq \Lambda $%
, one can obtain the corresponding trajectory $\upsilon $ as a sequence of
the counts of the jumps of $n_t(\Delta )$ in the time-space ${\bf R}%
_{+}\times \Lambda $.

Given an initial state vector in ${\cal H}_\infty $ of the form $\psi
_0=\eta \otimes \varphi _0$ with fixed $\varphi _0=f_0^\infty $, one can
define a nonunitary stochastic evolution $\eta \mapsto T(t,\upsilon )\eta $
by setting 
\[
T(t,\upsilon )=\langle \mbox{\boldmath $\lambda$}|U(t,\tau )\varphi
_0\,,\quad \upsilon =(\tau ,\mbox{\boldmath $\lambda$}), 
\]
which reproduces the unitary evolution $U(t,\tau )=U_0(t)V_t^{\dagger }(\tau
)$ defined by (\ref{eq:bhaml2.2}). This can also be written as $T(t,\upsilon
)=U(t)F_t^{\dagger }(\upsilon )$, since $U_0(t)=U(t)\otimes {\bf 1}^{\otimes
\infty }$ commutes with the (generalized) eigen-bras $\langle %
\mbox{\boldmath $\lambda$}|=\langle \lambda _1,\lambda _2,\ldots |$ of the
bubble coordinates $(p_1,p_2,\ldots ):\langle \mbox{\boldmath $\lambda$}%
|U_0(t)=U(t)\langle \mbox{\boldmath $\lambda$}|$. The reduction
transformations $F_t(\upsilon )$, $\upsilon \in \Upsilon _\infty $, are
given by the chronological products 
\begin{equation}
F_t(\upsilon )=G_t^{\dagger }(y_1)G_t^{\dagger }(y_2)\ldots \equiv
\prod_{y\in \upsilon }^{\rightarrow }G_t^{\dagger }(y)  \label{eq:bhaml2.7}
\end{equation}
of $G_t(t_n,\lambda _n)=U^{\dagger }(t_n)G(\lambda _n)U(t_n)$ for $t_n<t$,
where $G(\lambda )=\langle \lambda |Sf_0$, owing to the product form (\ref%
{eq:bhaml2.3}) of the unitary transformations $V_t(\tau )$, $\tau \in \Gamma 
$, and $\langle \mbox{\boldmath $\lambda$}|=\otimes _{k=1}^\infty \langle
\lambda _k|$, $\varphi _0(\mbox{\boldmath $\lambda$})=\prod_{k=1}^\infty
f_0(\lambda _k)$, $\langle \mbox{\boldmath $\lambda$}|\varphi _0\equiv 1$
for $\mbox{\boldmath $\lambda$}\in \Lambda ^\infty $. The stochastic
operator (\ref{eq:bhaml2.7}) defined by the single--point reductions 
\begin{equation}
G_t(r,\lambda )=\langle \lambda |S_t(r)f_0=\left\{ 
\begin{array}{ll}
G(r,\lambda )\,, & r<t \\ 
I\,, & r\geq t%
\end{array}
\right.  \label{eq:bhaml2.8}
\end{equation}
is normalized with respect to the initial probability ${\rm P}_0({\rm d}%
\upsilon )$, $\Pi _{\Upsilon _\infty }[I](t)=I$, where 
\begin{equation}
\Pi _A[X](t)=\int_AT^{\dagger }(t,\upsilon )XT(t,\upsilon ){\rm P}_0({\rm d}%
\upsilon )=\int_AF_t(\upsilon )XF_t^{\dagger }(\upsilon ){\rm P}_0({\rm d}%
\upsilon )\,,  \label{eq:bhaml2.9}
\end{equation}
is a continual operational-valued measure \cite{bib:bhaml3}-- \cite%
{bib:ref33} defined on measurable sets $A\subseteq \Upsilon _\infty $ of the
point trajectories $\upsilon _t=\{(r,\lambda )\in \upsilon |r<t\}$ given by
the operations $\Phi _t(\upsilon ):X\mapsto F_t(\upsilon )XF_t^{\dagger
}(\upsilon )$ for particle operators $X:{\cal H}\rightarrow {\cal H}$.

The positive operator-valued measure $\Pi _t(A)=\Pi _A[I](t)$ gives the
statistics 
\[
{\rm P}_t(A)=\int_A\Vert \langle \mbox{\boldmath $\lambda$}|U(t,\tau )\psi
_0\Vert ^2\mu _0^\infty ({\rm d}\mbox{\boldmath $\lambda$})\pi _0({\rm d}%
\tau _t) 
\]
for the continual observation with respect to an arbitrary initial
wave-function $\psi _0=\eta \otimes f_0^\infty $, $\eta \in {\cal H}$ in the
form 
\[
{\rm P}_t(A)=\langle \eta ,\Pi _t(A)\eta \rangle \,. 
\]
The output probability measure ${\rm P}(A)$, $A\subseteq \Upsilon _\infty $,
is defined by the marginales ${\rm P}_t(A)$, $A\subseteq \Upsilon _t$, as $%
t\rightarrow \infty $.

The reduced wave function $\chi (t,\upsilon )=T(t,\upsilon )\eta $ is
normalized 
\[
\Vert \chi (t)\Vert ^2=\int \Vert \chi (t,\upsilon )\Vert ^2{\rm P}_0({\rm d}%
\upsilon )=1 
\]
as a stochastic vector process $\chi (t):\Upsilon _\infty \rightarrow {\cal H%
}$ with respect to the initial probability distribution ${\rm P}_0$ on $%
\Upsilon _\infty $.

It satisfies the stochastic wave equation 
\begin{equation}
{\rm d}\chi (t)+\frac{{\rm i}}{\hbar }\,H\chi (t){\rm d}t={\rm d}%
n_{t}[G-I]\chi (t)\,,\quad \chi (0)=\eta \,,  \label{eq:bhaml2.10}
\end{equation}%
expressed in terms of the random differential ${\rm d}n_{t}[G-I](\upsilon
)=(G(\lambda _{n_{t}}(\upsilon ))-1){\rm d}n_{t}(\upsilon )$, $%
n_{t}(\upsilon )=n_{t}(\tau )$ for the point distribution $%
n_{t}[L]=\int_{\Lambda }L(\lambda )n_{t}({\rm d}\lambda )=n[L_{t}]$ over $%
\lambda \in \Lambda $ defined in (\ref{eq:bhaml2.6}) for $L_{t}(r,\lambda
)=1_{t}(r)L(\lambda )$.

To prove Eq.(\ref{eq:bhaml2.10}), discovered for the first time in \cite%
{bib:ref19}, we rewrite it in the following integral form 
\[
\chi (t)=e^{-{\rm i}Ht/\hbar }\biggl(\eta +\int_{0}^{t}\int_{\Lambda }e^{%
{\rm i}Hr/\hbar }(G(\lambda )-I)\chi (r){\rm d}n_{r}({\rm d}\lambda )\biggr)%
, 
\]%
given for each $\upsilon \in \Upsilon _{\infty }$ by the finite sum 
\[
n[1_{t}U^{\dagger }(r)(G-I)\chi ](\upsilon )=\sum_{(r,\lambda )\in \upsilon
}^{r<t}U^{\dagger }(r)(G(\lambda )-I)\chi (r)\,. 
\]%
We express the solution to this equation in the form $\chi (t,\upsilon
)=U(t)F_{t}^{\dagger }(\upsilon )\eta $ via the solution (\ref{eq:bhaml2.7})
to the recursion equation 
\[
F_{t}(\upsilon )=I+\sum_{(r,\lambda )\in \upsilon }^{r<t}F_{r}(\upsilon
)(G(r,\lambda )-I)\,,\quad F_{0}(\upsilon )=I, 
\]%
with $G(t,\lambda )=U^{\dagger }(t)G(\lambda )U(t)$, as was done for the
unitary case.

Let us also write on $F$ the nonlinear equation 
\[
{\rm d}\chi _{\upsilon }(t)+\frac{{\rm i}}{\hbar }\,H\chi _{\upsilon }(t)%
{\rm d}t=(G_{\upsilon }(t)-I)\chi _{\upsilon }(t){\rm d}n_{t}(\upsilon
)\,,\quad \chi _{\upsilon }(0)=\eta , 
\]%
with $G_{\upsilon }(t)=G(\lambda _{n_{t}(\upsilon )})/\Vert G(\lambda
_{n_{t}(\upsilon )})\chi _{\upsilon }(t)\Vert $. Its solutions define the
normalized reduction $\chi _{\upsilon }(t)=\chi (t,\upsilon )/\Vert \chi
(t,\upsilon )\Vert $ for continual counting measurements as a stochastic
vector process $\chi _{\upsilon }(t)\in {\cal H}$ with respect to the output
probability measure ${\rm P}$ of the point process $t\mapsto \upsilon _{t}$.
This can easily be obtained, as in \cite{bib:ref21} by applying the Ito
multiplication table 
\[
({\rm d}t)^{2}=0\,,\quad {\rm d}t{\rm d}n_{t}=0={\rm d}n_{t}{\rm d}t\,,\quad
({\rm d}n_{t})^{2}={\rm d}n_{t}\,. 
\]

\section{Mixing Reduction for Many Identical Particles}

\typeout{A Mixing Reduction for Many Identical Particles} \label{sec:bhaml3} %
\setcounter{equation}{0}

We now consider $M$ identical particles (atoms) interacting independently
with the bubbles (photons) in accordance with the scattering term in the
Hamiltonian (\ref{eq:bhaml2.1}). The spontaneous process of scatterings is
described by the time-ordered sequences of pairs $(k_n,t_n)$, $%
t_1<t_2<\ldots $, where $k_n\in \{1,\ldots ,M\}$ is the number of the
particle labeled by the scattering number $n\in {\bf N}$ at time instant $%
t_n>0$. We have excluded possibility of two or more scatterings of the
bubbles at the same instant of time, as was done for a single particle in
Sec.\ref{sec:bhaml2}. The sequence $(k_1,t_1)$, $(k_2,t_2),\ldots $ of the
scatterings can be represented by the occupational subsets $\tau _k=\{t_n\in
\tau |k_n=k\}$ of the time set $\tau =\{t_1,t_2,\ldots \}$, which are
disjoint, $\tau _k\cap \tau _l=\emptyset $ if $k\neq l$ since the
scatterings for different particles are independent. We shall consider the $%
M $-tuples $\tau _{\bullet }=(\tau _1,\ldots ,\tau _M)$ of these countable
subsets $\tau _k\subset {\bf R}_{+}$ as elements $\tau _{\bullet }\in \Gamma
_\infty ^M$ of the Cartesian $M$- product of $\Gamma _\infty $, given by the
partition $\tau =\sqcup \tau _k:\,=\cup \tau _k$, $\tau _k\cap \tau
_l=\emptyset $ if $k\neq l$ of a $\tau \in \Gamma _\infty $.

The Hamiltonian of the interacting particles for fixed $\tau _{\bullet
}=\{(k_1,t_1),\newline
(k_2,t_2),\ldots \}$ reads 
\begin{equation}
H(t,\tau _{\bullet })=H_0^M-\kappa \sum_{n=1}^\infty R(k_n))\otimes \delta
(t-t_n)Q(n)\,.  \label{eq:bhaml3.1}
\end{equation}
Here $H_0^M=H^M\otimes {\bf 1}$ is the Hamiltonian of the particles
describing the time evolution on the intervals between the scatterings with
the bubbles: 
\[
H^M=\sum_{k=1}^MH(k)+\sum_{k=1}^M\sum_{l>k}^MW(k,l)\,, 
\]
where $H(k)=I^{\otimes (k-1)}\otimes H\otimes I^{\otimes (M-k)}$ is the
Hamiltonian of the $k$th particle and $W(k,l)$ is the interaction potential
in ${\cal H}^{\otimes M}$ of the $k$th and $l$th particle, $1\leq k<l\leq M$.

Let ${\cal H}^{M}$ denote the $M$-particle Hilbert space, which is an
invariant subspace ${\cal H}^{M}\subseteq {\cal H}^{\otimes M}$ of symmetric
(bosons) or antisymmetric (fermions) $M$-tensors $\eta ^{M}\in {\cal H}%
^{\otimes M}$ generated by the product-vectors $\otimes _{k=1}^{M}\eta
_{k}\in {\cal H}^{\otimes M}$ with $\eta _{k}\in {\cal H}$. The
correspondent Ito-Schr\"{o}dinger equation for the stochastic state vector $%
\psi ^{M}(t):\tau _{\bullet }\mapsto \psi (t,\tau _{\bullet })$ with values $%
\psi (t,\tau _{\bullet })\in {\cal H}_{\infty }^{M}$ in the product space $%
{\cal H}_{\infty }^{M}={\cal H}^{M}\otimes {\cal E}$ of the $M$-particle
space ${\cal H}^{M}$ by ${\cal E}=\lim_{n\rightarrow \infty }L^{2}(\Lambda
^{n})$ reads 
\begin{equation}
{\rm d}\psi ^{M}(t)+\frac{{\rm i}}{\hbar }\,H_{0}^{M}\psi ^{M}(t){\rm d}%
t=(S(k_{t},n_{t})-I)\psi ^{M}(t){\rm d}n_{t}\,.  \label{eq:bhaml3.2}
\end{equation}%
Here $S(k,n)=\exp \{\frac{{\rm i}}{\hbar }\,\kappa R(k)\otimes Q(n)\}$, $%
R(k)=I^{\otimes (k-1)}\otimes R\otimes I^{\otimes (M-k)}$, 
\[
k_{t}(\tau _{\bullet })=\sum_{k=1}^{M}k1_{\tau _{k}}(t)\,,\quad \mbox{where}%
\;1_{\tau }(t)=\left\{ 
\begin{array}{ll}
1\,, & t\in \tau \\ 
0\,, & t\notin \tau%
\end{array}%
\right. 
\]%
is the random number $k_{t}:\Gamma _{\infty }^{M}\rightarrow \{1,\ldots ,M\}$%
, labeling a particle by $k$ at any instant $t\in \tau _{k}$ of its
collision with a bubble labeled by $n_{t}(\tau )=\sum_{k=1}^{M}n_{k,t}=|\tau
\cap \lbrack 0,t)|$, where $n_{k,t}=|\tau _{k}\cap \lbrack 0,t)|$, $\tau
=\cup \tau _{k}$.

The solutions $\psi (t,\tau _{\bullet })=U(t,\tau _{\bullet })\psi _{0}^{M}$%
, $\psi _{0}^{M}\in {\cal H}_{\infty }^{M}$, of equation (\ref{eq:bhaml3.2})
can be written as $U(t,\tau _{\bullet })=U_{0}^{M}(t)V_{t}^{\dagger }(\tau
_{\bullet })$ in terms of the finite chronological product 
\begin{equation}
V_{t}(\tau _{\bullet })=S_{t}^{\dagger }(k_{1},t_{1})S_{t}^{\dagger
}(k_{2},t_{2})\ldots ,\quad \tau _{\bullet }\in \Gamma _{\infty }^{M}\,,
\label{eq:bhaml3.3}
\end{equation}%
where $S_{t}(k,t_{n})=I$ if $t_{n}\geq t$, $S_{t}(k,t_{n})=U_{0}^{M\dagger
}(t_{n})S(k,n)U_{0}^{M}(t_{n})$ if $t_{n}<t$, and $U_{0}^{M}(t)=\exp \{-%
\frac{{\rm i}}{\hbar }\,H_{0}^{M}(t)\}$. The proof is exactly the same as
for the case of a single particle $M=1$.

Let $\omega =(w_1,w_2,\ldots )$ denote a chronologically ordered sequence of
triples $w_n=(k_n,t_n,\lambda _n)$ and $\Omega $ the space of all such
sequences with $\{t_1,t_2,\ldots \}\in \Gamma _\infty $. Every sequence $%
\omega \in \Omega $ can be represented as a pair $\omega =(\tau _{\bullet },%
\mbox{\boldmath $\lambda$})$, where $\tau _{\bullet }=(\tau _1,\ldots ,\tau
_M)$ is a partition of the corresponding sequence $\tau =\{t_1,t_2,\ldots \}$
and $\mbox{\boldmath $\lambda$}=(\lambda _1,\lambda _2,\ldots )$, so that $%
\Omega $ can be identified with the product $\Gamma _\infty ^M\times
\Upsilon _\infty $. The space $\Omega $ is equipped with the probability
measure ${\rm P}_0({\rm d}\omega )=\pi _0({\rm d}\tau _{\bullet })\mu
_0^\infty ({\rm d}\mbox{\boldmath $\lambda$})$, where $\pi _0({\rm d}\tau
_1,\ldots ,{\rm d}\tau _M)=\prod_{k=1}^M\pi _0({\rm d}\tau _k)$ is the
product of the identical Poisson measures (\ref{eq:bhaml2.5}), in accordance
with the independence of the spontaneous interactions of each particle with
the bubbles.

Given an initial state vector $\psi ^{M}=\eta ^{M}\otimes \varphi _{0}$,
where $\varphi _{0}=f_{0}^{\infty }$, one can easily prove that the
nonunitary stochastic evolution 
\[
T(t,\omega )=\langle \mbox{\boldmath $\lambda$}|U(t,\tau _{\bullet })\varphi
_{0}\,,\quad \omega =(\tau _{\bullet },\mbox{\boldmath $\lambda$}) 
\]%
is also a finite chronological product 
\[
T(t,\omega )=U^{M}(t)F_{t}^{\dagger }(\omega )\,,\quad F_{t}(\omega
):=G_{t}^{\dagger }(w_{1})G_{t}^{\dagger }(w_{2})\ldots \,. 
\]%
Here $U^{M}(t)=\exp \{-\frac{{\rm i}}{\hbar }\,H^{M}t\}$, $%
G_{t}(k,t_{n},\lambda )=I$ for $t_{n}\geq t$, and $G_{t}(k,t_{n},\lambda
)=U^{M\dagger }(t_{n})G(k,\lambda )U^{M}(t_{n})$, $t_{n}<t$, is defined by
the reduced scattering operator 
\begin{equation}
G(k,\lambda )=I^{\otimes (k-1)}\otimes G(\lambda )\otimes I^{\otimes
(M-k)}\,,\quad G(\lambda )=\langle \lambda |Sf_{0}  \label{eq:bhaml3.4}
\end{equation}%
for $f_{0}\in L^{2}(\Lambda )$, $\Vert f_{0}\Vert =1$, applied to the $k$th
particle only in ${\cal H}^{M}$. The stochastic operator $T(t)$ defines the
solutions $\chi ^{M}(t,\omega )=T(t,\omega )\eta ^{M}$ to the Ito
differential equation 
\[
{\rm d}\chi ^{M}(t)+\frac{{\rm i}}{\hbar }H^{M}\chi ^{M}(t){\rm d}%
t=\sum_{k=1}^{M}{\rm d}n_{k,t}[G(k)-I]\chi ^{M}(t)\,,\quad \chi ^{M}(0)=\eta
^{M} 
\]%
for the stochastic vector states $\chi ^{M}(t):\Omega \rightarrow {\cal H}%
^{M}$ of the $M$-particle system, corresponding to an initial $\eta ^{M}\in 
{\cal H}^{M}$. The right--hand side of this equation is written as the
forward increment of the point integral 
\[
n[L_{t}]=\sum_{k=1}^{M}\int_{0}^{t}\int_{\Lambda }L(k,r,\lambda ){\rm d}%
n_{k,r}({\rm d}\lambda ) 
\]%
given by the stochastic distribution $n[L](\omega )=\sum_{w\in \omega }L(w)$
for $L_{t}(k,r,\lambda )=1_{t}(r)L(k,r,\lambda )$. The vector $\chi
^{M}(t,\omega )\in {\cal H}^{M}$ as well as $\chi (t,\upsilon )$ in (\ref%
{eq:bhaml2.10}), is no longer normalized $(\Vert \chi ^{M}(t,\omega )\Vert
\neq 1)$ for an initial state-vector $\eta ^{M}\in {\cal H}^{M}$, $\Vert
\eta ^{M}\Vert =1$, but it is normalized with respect to the probability
measure ${\rm P}_{0}$ on $\Omega $ in the mean square sense. But, in
contrast to $\chi (t,\upsilon )$, $\chi ^{M}(t,\omega )$ is not yet the
reduced description of the $M$-particle system under the observation of the
scattering process $\upsilon _{t}=\{(r,\lambda )\in \upsilon |r\leq t\}$,
given by the registration of the pointer positions $\lambda _{n}\in \Lambda $
at random time instants $t_{n}$.

The reduced dynamics corresponding to the observation is described by a
stochastic operational process $X\mapsto \Theta [X](t)$, 
\begin{equation}  \label{eq:bhaml3.5}
\Theta [X](t,\upsilon )=\Phi _t[U^{M\dagger }(t)XU^M(t)](\upsilon )
\end{equation}
for the $M$-particle operators $X:{\cal H}^M\rightarrow {\cal H}^M$ given by
the conditional expectation 
\begin{equation}  \label{eq:bhaml3.6}
\Phi _t[X](\tau ,\mbox{\boldmath $\lambda$})=\frac 1{M^{|\tau
_t|}}\,\sum_{\sqcup \sigma _k=\tau }F_t(\sigma _{\bullet },%
\mbox{\boldmath
$\lambda$})XF_t^{\dagger }(\sigma _{\bullet },\mbox{\boldmath $\lambda$})\,
\end{equation}
where the sum is taken over all partitions $\sigma _{\bullet }=(\sigma
_1,\ldots ,\sigma _M)$ of a finite subset $\tau _t=\tau \cap [0,t)$. This
averaging is due to the impossibility to detect the individuality of the
identical particles producing the indistinguishable effects on the bubbles
by measuring the scatterings of the bubbles.

To prove Eq. (\ref{eq:bhaml3.6}), we need to compare the correlations of $%
F_t(\omega )XF_t^{\dagger }(\omega )$ and of an arbitrary functional $%
g(\upsilon _t)$ of the observable point process $\upsilon _t$ with the
correlations of (\ref{eq:bhaml3.6}) and of $g(\upsilon _t)$. But by applying
the well known formula \cite{bib:DM3} one can easily find 
\[
\int_{\Gamma _t^M}x(\sigma _1,\ldots ,\sigma _M)\prod_{k=1}^M{\rm d}\sigma
_k=\int_{\Gamma _t}\sum_{\sigma _k:\sqcup \sigma _k=\sigma }x(\sigma
_1,\ldots ,\sigma _M){\rm d}\sigma 
\]
for the multiple point integration that these correlations with respect to
the probability measure ${\rm P}_0$ on $\Omega $ given by the Poisson law (%
\ref{eq:bhaml2.5}) simply coincide: 
\begin{eqnarray*}
\lefteqn{\int_{\Gamma_t^M}\pi_t({\rm d}\sigma_\bullet)\int_{\Lambda^\infty}
g(\sqcup_{k=1}^M\sigma_k,\bm\lambda)F(\sigma_\bullet,\bm\lambda)X
F_t^\dagger(\sigma_\bullet,\bm\lambda)\mu^\infty_0({\rm d}\bm\lambda) } \\
&&=\int_{\Gamma_t^M}\langle g(\sqcup^M_{k=1}\sigma_k),
X(\sigma_1,\ldots,\sigma_M)\rangle_0 \prod_{k=1}^Me^{-\nu t}\nu^{|\sigma_k|}%
{\rm d}\sigma_k \\
&&=\int_{\Gamma_t}\langle g(\sigma), \sum_{\sigma_k:\sqcup\sigma_k=\sigma}
X(\sigma_1,\ldots,\sigma_M)\rangle_0 e^{-M\nu t}\nu^{|\sigma|}{\rm d}\sigma
\\
&&=\int_{\Gamma_t}\pi^M_t({\rm d}\sigma)\int_{\Lambda^\infty} g(\sigma,%
\mbox{\boldmath $\lambda$})\frac
1{M^{|\sigma|}}\,\sum_{\tau_k:\cup\tau_k=\sigma}
F(\omega)XF^\dagger(\omega)\mu^\infty_0({\rm d}\mbox{\boldmath $\lambda$})\,.
\end{eqnarray*}
Here $\langle .,.\rangle _0$ is the abbreviation for the inner product in $%
{\cal E}$ of the test function $\upsilon \mapsto g(\tau ,%
\mbox{\boldmath
$\lambda$})$ with fixed $\tau \in \Gamma _t$ and the operator function $%
X_t(\tau _{\bullet },\mbox{\boldmath $\lambda$})=F_t(\tau _{\bullet },%
\mbox{\boldmath $\lambda$})XF_t^{\dagger }(\tau _{\bullet },%
\mbox{\boldmath
$\lambda$})$ with fixed $\tau _{\bullet }\in \Gamma _t^M$. The probability
measure 
\begin{equation}  \label{eq:bhaml3.7}
\pi _0^M({\rm d}\tau _t)=\sum_{\sqcup \sigma _k=\tau _t}\prod_{k=1}^M\pi _0(%
{\rm d}\sigma _k)=e^{-M\nu _t}|M\nu |^{|\tau _t|}{\rm d}\tau _t
\end{equation}
on $\Gamma _t$ has the intensity $M\nu $. It is induced by the measure $\pi
_0({\rm d}\tau _{\bullet })$ on $\Gamma _\infty ^M$ with respect to the
particle identification map $\tau _{\bullet }\in \Gamma _\infty ^M\mapsto
\tau =\cup _{k=1}\tau _k$, defining the observable data $\upsilon =(\tau ,%
\mbox{\boldmath $\lambda$})$ by the stochastic map $\omega =(\tau _{\bullet
},\mbox{\boldmath $\lambda$})\mapsto \upsilon \in \Upsilon _\infty $ on $%
\omega \in \Omega $.

By the coincidence of the correlations proved above, the stochastic operator
(\ref{eq:bhaml3.5}) is indeed the conditional expectation of the stochastic
operators $X_t(\omega )$ with respect to the observable process $\upsilon _t$%
.

In contrast to the pure operations $X\mapsto X_t(\omega )$ the reduction
operation $X\mapsto \Phi _t[X](\upsilon )$, preserves the symmetry of the $M$%
-particle operators $X$ with respect to particle permutation. It is the
least mixing operation which preserves the indistinguishability of the
particles respectively to the observations of the bubble scatterings,
corresponding to the complete nondemolition measurement of the particles.

Indeed, the reduction operation (\ref{eq:bhaml3.6}) can be simply written as
the finite iteration 
\begin{eqnarray*}
\frac 1{M^n}\sum_{k_1,\ldots,k_n=1}^MG^\dagger_t(k_1,y_1)\ldots
G^\dagger_t(k_n,y_n)XG_t(k_n,y_n)\ldots G_t(k_1,y_1) \\
=\Psi_t[\ldots\Psi_t[X](y_1)\ldots](y_n)=\Phi_t[X](y_1,\ldots,y_n)
\end{eqnarray*}
with $n=|\tau _t|$ single mixing reductions 
\begin{equation}  \label{eq:bhaml3.8}
\Psi [X](y)=\frac 1M\sum_{k=1}^MG^{\dagger }(k,y)XG(k,y)\,.
\end{equation}
Given as the arithmetric mean value of the permutations for the pure
operations $X\mapsto G_t^{\dagger }(k,y)XG_t(k,y)$, corresponding to the
identical operators (\ref{eq:bhaml3.4}), the reductions $X\mapsto \Psi
_t[X](y)$ are permutationally symmetric, and are not mixing only if the pure
operations if they do not break this symmetry.

The derived mixing property of the reduced stochastic dynamics $t\mapsto
\rho \lbrack X](t,\upsilon )$ obtained above for the corresponding
statistical states 
\begin{equation}
\rho ^{M}[X](t,\upsilon )=\langle \eta ^{M},\Theta \lbrack X](t,\upsilon
)\eta ^{M}\rangle ={\rm Tr}\{X\varrho ^{M}(t,\upsilon )\}
\label{eq:bhaml3.9}
\end{equation}%
gives an increase in the entropy 
\[
\sigma ^{M}(t,\upsilon )=-{\rm Tr}\{\varrho ^{M}(t,\upsilon )\ln \varrho
^{M}(t,\upsilon )\} 
\]%
for an ensemble of identical particles even under the condition of complete
nondemolition observation. According to (\ref{eq:bhaml3.9}), the reduced
density operators $\varrho ^{M}(t,\upsilon )$ for the system of $M$
identical particles gives the probability density 
\[
{\rm p}^{M}(t,\upsilon _{t})={\rm Tr}\{\varrho ^{M}(t,\upsilon _{t})\}={\rm P%
}^{M}({\rm d}\upsilon _{t})/{\rm P}_{0}^{M}({\rm d}\upsilon _{t}) 
\]%
of the output process $\upsilon _{t}$. Here ${\rm P}_{0}^{M}=\pi ^{M}\otimes
\mu _{0}^{\infty }$ is the probability measure on $\Upsilon _{\infty
}=\Gamma _{\infty }\times \Lambda ^{\infty }$ defined by the Poisson measure
(\ref{eq:bhaml3.7}). This means that the a posteriori density operator $%
\varrho ^{M}(t)$ is defined as a stochastic positive trace class operator
normalized in the mean sense 
\[
\Vert \varrho ^{M}(t)\Vert _{1}=\int {\rm Tr}\{\varrho ^{M}(t,\upsilon )\}%
{\rm P}_{0}^{M}({\rm d}\upsilon )=1\,. 
\]%
The density $\varrho ^{M}(t)$ satisfies the stochastic operator equation 
\begin{equation}
{\rm d}\varrho ^{M}(t)+\frac{{\rm i}}{\hbar }\,[H,\varrho ^{M}(t)]{\rm d}t=%
{\rm d}n_{t}\biggl[\frac{1}{M}\,\sum_{k=1}^{M}G(k)\varrho ^{M}(t)G^{\dagger
}(k)-\varrho ^{M}(t)\biggr],  \label{eq:bhaml3.10}
\end{equation}%
which has a unique solution for every initial condition $\varrho
^{M}(0,\upsilon )=\varrho _{0}^{M}$ given by the density operator $\varrho
_{0}^{M}$ for the $M$-particle states $\varrho _{0}^{M}[X]={\rm Tr}%
\{X\varrho _{0}^{M}\}$.

Let derive the differential equation (\ref{eq:bhaml3.10}) in the equivalent
integral form 
\[
\rho ^{M}[X](t)=\rho ^{M}[X(t)](t)+\int_{0}^{t}\int_{\Lambda }\rho
^{M}[(\Psi \lbrack X(t-r)](\lambda ))-X(t-r)](r){\rm d}n_{t}({\rm d}\lambda
) 
\]%
for an $\rho _{0}^{M}[X]=\langle \eta ^{M},X\eta ^{M}\rangle $, $\eta
^{M}\in {\cal H}^{M}$, where $X(t)=U^{M}(t)^{\dagger }XU^{M}(t)$, 
\[
U^{M}(t)=e^{-{\rm i}H^{M}t/\hbar }\,,\quad \Psi \lbrack X](\lambda )=\frac{1%
}{M}\sum_{k=1}^{M}G^{\dagger }(k,\lambda )XG(k,\lambda )\,. 
\]%
Taking into account the fact that the stochastic integral (\ref{eq:bhaml2.6}%
) in this equation is simply a finite sum for every $\upsilon \in \Upsilon
_{\infty }$ and $t$, one can write it as recursive operator equation 
\[
\Phi _{t}[X](\upsilon )=X+\sum_{(r,y)\in \upsilon }^{r<t}\Phi _{r}[\Psi
\lbrack X](r,\lambda )-X](\upsilon )\,, 
\]%
for a stochastic operation $\Phi _{t}$, defining the solutions to Eq. (\ref%
{eq:bhaml3.10}), as in (\ref{eq:bhaml3.9}), in terms of the composition (\ref%
{eq:bhaml3.5}).

But such recurrence has the unique solution $\Phi _t(\upsilon )=\Psi
_t(y_1)\circ \Psi _t(y_2)\circ \ldots $, defined for a $\upsilon =(\tau ,%
\mbox{\boldmath $\lambda$})\in \Upsilon _\infty $ as the chronological
composition of the maps $\Psi _t(r,\lambda ):X\mapsto \Psi [X](r,\lambda )$
if $r<t$ and $\Psi _t[X](r,\lambda )=X$ if $r\geq t$. This solution can be
found by the iterations 
\begin{eqnarray*}
\lefteqn{\Phi_t(\upsilon)={\rm I}+
\sum_{(r,\lambda)\in\upsilon}^{r<t}\Phi_r(\upsilon)\circ\Lambda(r,\lambda)}
\\
&&={\rm I}+\sum_{(r,\lambda)\in\upsilon}^{r<t}\biggl({\rm I}+
\sum_{(s,\lambda)\in\upsilon}^{s<r}\Phi_s(\upsilon)\circ\Lambda(s,\lambda)%
\biggr) \circ\Lambda(r,\lambda) \\
&&=\ldots=\sum_{\sigma\subseteq\upsilon_t}\Lambda(z_1)\circ\ldots\circ%
\Lambda(z_n)\,,
\end{eqnarray*}
where $\Lambda (y)=\Psi (y)-{\rm I}$, $\Phi _s(\upsilon )$ is the identical
map ${\rm I}:X\mapsto X$ if $s=t_1$ and $\sum_{\sigma \subseteq \upsilon
_t}\Lambda (z_1)\circ \ldots \circ \Lambda (z_n)=\Psi _t(y_1)\circ \Psi
_t(y_2)\circ \ldots $ in terms of $\sigma =\{z_1,\ldots ,z_n\}$, $%
z=(r,\lambda )$, $s_1<\ldots <s_n$, $n\leq n_t$.

Let us also write the nonlinear stochastic equation 
\[
{\rm d}\varrho _{\upsilon }^{M}(t)+\frac{{\rm i}}{\hbar }\,[H,\varrho
_{\upsilon }^{M}(t)]{\rm d}t=\varrho _{\upsilon }^{M}(t)\circ (\Psi
_{\upsilon }(t)-{\rm I}){\rm d}n_{t}(\upsilon )\,, 
\]%
where $\Psi _{\upsilon }(t)=\Psi (\lambda _{n_{t}(\upsilon )})/{\rm Tr}%
\{E(\lambda _{n_{t}(\upsilon )})\rho \}$, 
\[
\varrho \circ \Psi (\lambda )=\frac{1}{M}\,\sum_{k=1}^{M}G(k,\lambda
)\varrho G^{\dagger }(k,\lambda )\,,\quad E(\lambda )=\frac{1}{M}%
\,\sum_{k=1}^{M}G^{\dagger }(k,\lambda )G(k,\lambda )) 
\]%
for the normalized density operator 
\[
\varrho _{\upsilon }^{M}(t)=\varrho ^{M}(t,\upsilon )/{\rm p}^{M}(t,\upsilon
)\,. 
\]%
This describes the conditional expectations $\rho _{\upsilon }^{M}[X](t)=%
{\rm Tr}\{X\varrho _{\upsilon }^{M}(t)\}$ of the $M$-particle operators with
respect to the output probability measure 
\[
{\rm P}^{M}({\rm d}\upsilon )={\rm p}^{M}(t,\upsilon ){\rm P}_{0}^{M}({\rm d}%
\upsilon )\,, 
\]%
where ${\rm p}^{M}(t,\upsilon )={\rm Tr}\{\varrho ^{M}(t,\upsilon )\}$.

\section{Macroscopic and Continuous Reduction Limits}

\typeout{Macroscopic and Continuous Reduction Limits} \label{sec:bhaml4} %
\setcounter{equation}{0}

We now consider the mean field approximation of the measurement apparatus
fixing its total effect $\nu \kappa =\gamma $ given by the mean number $\nu $
of scattered bubbles (photons) per second and an interaction constant $%
\kappa $ coupling each bubble to a particle (atom) in the Hamiltonian (\ref%
{eq:bhaml2.1}). We look for the limits of the unitary and reduced evolutions
(\ref{eq:bhaml2.2}) and (\ref{eq:bhaml2.10}) as $\nu \rightarrow \infty $
and $\kappa \rightarrow 0$ such that $\gamma $ is a real constant. To
perform these limits we need the expansions 
\begin{eqnarray}
S(n) &=&I\otimes {\bf 1}+{\rm i}\frac{\kappa }{\hbar }\,R\otimes Q(n)-\biggl(%
\frac{1}{2}\,\biggr)\biggl(\frac{\kappa }{\hbar }\biggr)^{2}\,(R\otimes
Q(n))^{2}+\ldots  \label{eq:bhaml4.1} \\
G(\lambda ) &=&I-\kappa \frac{f_{0}^{\prime }(\lambda )}{f_{0}(\lambda )}\,R+%
\frac{1}{2}\,\kappa ^{2}R\frac{f_{0}^{\prime \prime }(\lambda )}{%
f_{0}(\lambda )}\,R+\ldots  \nonumber
\end{eqnarray}%
of the scattering operator $S(n)=\exp \{\frac{{\rm i}}{\hbar }\,\kappa
R\otimes Q(n)\}$ and the reduced operator $G(\lambda )=f_{0}(\lambda
I-\kappa R)/f_{0}(\lambda )$ with respect to the coupling constant $\kappa $%
. The first term of the expansion for $S(n)$ is disappearing in the r.h.s.
of Eq.(\ref{eq:bhaml2.2}) while the second and third terms are appearing as
the differentials of the operator-valued stochastic integrals 
\[
\hat{n}_{t}[Q]=\int_{0}^{t}Q(n_{r}){\rm d}n_{r}\,,\quad \hat{n}%
_{t}[Q^{2}]=\int_{0}^{t}Q(n_{r})^{2}{\rm d}n_{r}\,. 
\]%
The corresponding terms 
\[
n_{t}\biggl[\frac{f_{0}^{\prime }}{-f_{0}}\biggr]=\int_{0}^{t}\int_{\Lambda }%
\frac{f_{0}^{\prime }(\lambda )}{-f_{0}(\lambda )}\,{\rm d}n_{r}({\rm d}%
\lambda )\,,\quad n_{t}\biggl[\frac{f_{0}^{\prime \prime }}{f_{0}}\biggr]%
=\int_{0}^{t}\int_{\Lambda }\frac{f_{0}^{\prime \prime }(\lambda )}{%
f_{0}(\lambda )}\,{\rm d}n_{r}({\rm d}\lambda ) 
\]%
in the right--hand side. of eq.(\ref{eq:bhaml2.10}) can also be written as
the integrals $\hat{n}_{t}[L]=\int_{0}^{t}L(n_{r}){\rm d}n_{r}$ with values
in operator functions of $\tau \in \Gamma _{\infty }$ 
\begin{equation}
\hat{n}_{t}[L](\tau )=\sum_{n=1}^{n_{t}(\tau )}L(n)\,,\quad L(n)={\bf 1}%
^{\otimes (n-1)}\otimes L\otimes {\bf 1}^{\otimes \infty }\,,
\label{eq:bhaml4.2}
\end{equation}%
where $L=[l(\lambda )]$ is one of the multiplication operators 
\[
L^{\prime }=-[f_{0}^{\prime }(\lambda )/f_{0}(\lambda )]\,,\quad L^{\prime
\prime }=[f_{0}^{\prime \prime }(\lambda )/f_{0}(\lambda )]\,. 
\]%
Here $f_{0}^{\prime }(\lambda )$ denotes the derivative $\partial
f_{0}(\lambda )=\partial f_{0}(\lambda )/\partial \lambda $ and $f^{\prime
\prime }(\lambda )$ denotes $\partial \partial f(\lambda )$. The stochastic
integral $\hat{n}_{t}[L](\tau )$ corresponding to the pointwise
multiplication $L:f\mapsto lf$ of $f\in L^{2}(\Lambda )$ by a function $l$
of the bubble coordinate $\lambda $ acts in ${\cal E}$ as the multiplication
operator 
\[
\lbrack \hat{n}_{t}[L](\tau )\varphi ](\mbox{\boldmath $\lambda$}%
)=\sum_{n=1}^{n_{t}(\tau )}l(\lambda _{n})\varphi (\mbox{\boldmath $\lambda$}%
)\equiv n_{t}[l](\tau ,\mbox{\boldmath $\lambda$})\varphi (%
\mbox{\boldmath
$\lambda$})\,.\label{eq:bhaml4.3} 
\]

Hence, the main terms on the right--hand side of Eq.(\ref{eq:bhaml2.2}) and (%
\ref{eq:bhaml2.10}) for $\kappa \rightarrow 0$ are given by the renormalised
stochastic integrals 
\begin{equation}  \label{eq:bhaml4.4}
\hat \lambda (t)=\frac 1{\nu t}\,\int_0^tL(n_r){\rm d}n_r=\frac 1{\nu
t}\,\hat n_t[L]
\end{equation}
of the operator-valued stochastic functions $L(t,\tau )=L(n_t(\tau ))$ with
respect to the numerical process $n_t(\tau )$ that has the Poisson
probability distribution (\ref{eq:bhaml2.5}) on $\Gamma _\infty $.

To pass to the large number limit $\nu \rightarrow \infty $ in (\ref%
{eq:bhaml4.4}) for an arbitrary operator $L$ in $L^2(\Lambda )$, we need to
use the quantum stochastic representation \cite{bib:12} of the integral (\ref%
{eq:bhaml4.4}) in the Fock space ${\cal F}$ over $L^2({\bf R}_{+}\times
\Lambda )$. The space ${\cal F}$ can be defined as the $L^2(\Upsilon )$%
-space of all square integrable functions $\varphi :\Upsilon \rightarrow 
{\bf C}$, $\Vert \varphi \Vert ^2=\int_\Upsilon |\varphi (\upsilon
)|^2\lambda ({\rm d}\upsilon )<\infty $ of time ordered finite sequences $%
\upsilon =(y_1,\ldots ,y_n)$, $y=(t,\lambda )$ identified with subsets $%
\upsilon \subset {\bf R}_{+}\times \Lambda $ of cardinality $|\upsilon
|=0,1,2,\ldots $. The measure $\lambda ({\rm d}\upsilon )$ on the union $%
\Upsilon =\sum_{n=0}^\infty \Upsilon (n)$ of the disjoint subsets $\Upsilon
(n)=\{\upsilon \in \Upsilon :|\upsilon |=n\}$ is given as the sum 
\[
\lambda (A)=\sum_{n=0}^\infty \lambda (\Upsilon (n)\cap A) 
\]
of the product $\lambda ({\rm d}\upsilon )=\prod_{y\in \upsilon }{\rm d}y$
of measures ${\rm d}y={\rm d}t{\rm d}\lambda $ on ${\bf R}_{+}\times \Lambda 
$ such that 
\[
\Vert \varphi \Vert ^2=\sum_{n=0}^\infty \int \negthinspace \negthinspace
\int_{0\leq t_1<\ldots \leq t_n<\infty }|\varphi (y_1,\ldots
,y_n)|^2\prod_{i=1}^n{\rm d}y_i\,. 
\]
Let define $N_t[L]$ define the numerical integral in ${\cal F}$ by the
action of the operator (\ref{eq:bhaml4.2})in each Fock component ${\cal F}%
(\tau )=L^2(\Lambda ^{|\tau |})$ for all $\tau \in \Gamma $: 
\begin{equation}  \label{eq:bhaml4.5}
[N_t[L]\varphi ](\tau )=\sum_{n=1}^{n_t(\tau )}L(n)\varphi (\tau )=\hat
n_t[L](\tau )\varphi (\tau )\,.
\end{equation}
Here $\varphi (\tau )$ is the function $\varphi (\tau ,%
\mbox{\boldmath
$\lambda$})=\varphi (\upsilon )$ of $\mbox{\boldmath $\lambda$}\in \Lambda
^{|\tau )}$, corresponding to a $\varphi \in L^2(\Upsilon )$ with a fixed
time component of $\upsilon =(\tau ,\mbox{\boldmath $\lambda$})\in \Upsilon $%
.

In order to obtain the initial probability measure ${\rm P}_{0}({\rm d}%
\upsilon )=\pi ({\rm d}\tau )\mu _{0}^{\infty }({\rm d}%
\mbox{\boldmath
$\lambda$})$ on $\Upsilon _{\infty }$ induced by an initial Fock vector $%
\varphi _{0}\in L^{2}(\Upsilon )$, we need an isomorphic transformation of (%
\ref{eq:bhaml4.5}) 
\begin{equation}
\hat{N}_{t}[L]=N_{t}[L]+\sqrt{\nu }(A_{t}[f_{0}^{\dagger }L]+A_{t}^{\dagger
}[Lf_{0}])+\nu tf_{0}^{\dagger }Lf_{0}  \label{eq:bhaml4.6}
\end{equation}%
which can be locally performed by a unitary transformation 
\[
\hat{N}_{t}[L]=U_{s}^{\dagger }N_{t}[L]U_{s}\,,\quad U_{s}=\exp \{\sqrt{\nu }%
(A_{s}^{\dagger }[f_{0}]-A_{s}[f_{0}^{\dagger }])\} 
\]%
for every $t<s$. Here $A_{t}^{\dagger }[f]$ and $A_{t}[f^{\dagger }]$ are
the creation and annihilation integrals of $f\in L^{2}(\Lambda )$, $%
f^{\dagger }\in L^{2}(\Lambda )^{\ast }$, given by the operators 
\begin{eqnarray*}
\Big[A_{t}^{\dagger }[f]\varphi \Big](\upsilon ) &=&\sum_{y\in \upsilon
_{t}}f(\lambda )\varphi (\upsilon \backslash y) \\
\Big[A_{t}[f^{\dagger }]\varphi \Big](\upsilon ) &=&\int_{([0,t)\times
\Lambda }f(\lambda )^{\ast }\varphi (\upsilon \sqcup y){\rm d}y
\end{eqnarray*}%
in the Fock space $L^{2}(\Upsilon )$, where $\upsilon \backslash y$ means
the sequence $\upsilon \in \Upsilon $ with deleted $y=(r,\lambda )$, $r<t$,
and $\upsilon \sqcup y$ means the sequence $\upsilon \in \Upsilon $ with an
additional element $y\notin \upsilon $. The characteristic functional of the
stochastic operators $\hat{n}_{t}[L]$ with respect to the initial
state-vector $f_{0}^{\infty }\in {\cal E}$ and the Poisson probability
measure (\ref{eq:bhaml2.5}) is now given simply by the vacuum expectation 
\[
\int_{\Gamma ^{\infty }}(f^{\infty },e^{{\rm i}\hat{n}_{t}[L]}f^{\infty
})\pi _{0}({\rm d}\tau )=\langle \delta _{\phi },e^{{\rm i}\hat{N}%
_{t}[L]}\delta _{\phi }\rangle \,, 
\]%
where $\delta _{\phi }(\upsilon )=1$ if $\upsilon =\emptyset $; otherwise, $%
\delta _{\phi }(\upsilon )=0$.

The corresponding representation $\hat{l}(t)=\frac{1}{\nu t}\,\hat{N}_{t}[L]$
for (\ref{eq:bhaml4.4}) helps us immediately obtain the quantum large number
limit 
\[
\lim_{\nu \rightarrow \infty }\frac{1}{\nu t}\,\hat{N}_{t}[L]=f_{0}^{\dagger
}Lf_{0}\hat{1} 
\]%
as the mean value $l_{0}=(f_{0},Lf_{0})\equiv f_{0}^{\dagger }Lf_{0}$ of a
single-bubble operator with respect to an initial wave packet $f_{0}\in
L^{2}(\Lambda )$. This gives the following macroscopic limit 
\[
{\rm d}\psi (t)+\frac{{\rm i}}{\hbar }\,H_{0}\psi (t){\rm d}t=\frac{{\rm i}}{%
\hbar }\,\gamma (R\otimes q_{0}\hat{1})\psi (t){\rm d}t 
\]%
of the generalized Schr\"{o}dinger equation (\ref{eq:bhaml2.2}) which turns
out to be a nonsingular one with an additional potential $-\gamma q_{0}R$
corresponding to the mean momentum $q_{0}=(f_{0},Qf_{0})$ of a bubble in the
initial state $f_{0}$. As one could expect, the mean field dynamics
preserves the product structure $\psi (t,\upsilon )=\eta (t)\varphi
_{0}(\upsilon )$ of an initial product-vector $\psi _{0}=\eta \otimes
\varphi _{0}$ being trivial on the Fock component $\varphi _{0}\in {\cal F}$
because of $H_{0}=H\otimes \hat{1}$. But unexpectedly (compare with \cite%
{bib:bel7}) the macroscopic limit 
\begin{equation}
{\rm d}\chi (t)+\frac{{\rm i}}{\hbar }\,H\chi (t){\rm d}t=\frac{{\rm i}}{%
\hbar }\,\gamma q_{0}R\chi (t){\rm d}t  \label{eq:bhaml4.7}
\end{equation}%
of the reduction equation (\ref{eq:bhaml2.10}) corresponds to the same
unitary dynamics $\eta (t)=U(t)\eta =\chi (t)$ of the particle state-vector
if $\chi (0)=\eta $, because of 
\[
\frac{1}{\nu t}\,n_{t}\biggl[\frac{f_{0}^{\prime }}{-f_{0}}\biggr]%
\,\rightarrow (f_{0},L^{\prime }f_{0})=\int f_{0}(\lambda )^{\ast
}f_{0}^{\prime }(\lambda ){\rm d}\lambda =\frac{{\rm i}}{\hbar }\,q_{0}\,. 
\]%
The macroscopic limits for the $M$-particle system also give essentially the
same continuous unitary evolutions in the large space ${\cal H}^{M}\otimes 
{\cal F}$ and in the reduced space ${\cal H}^{M}$. To get this
correspondence, one has only to replace the measure (\ref{eq:bhaml2.5}) on $%
\Gamma _{\infty }$ for Eq.(\ref{eq:bhaml3.2}) by the product measure $\pi
_{0}^{\otimes M}$ on $\Gamma _{\infty }^{M}$ and for Eq.(\ref{eq:bhaml3.10})
by the induced measure (\ref{eq:bhaml3.7}) on $\Gamma _{\infty }$, taking $%
M\nu $ instead of $\nu $ in (\ref{eq:bhaml4.6}). This means that the mixing
property of the reduced equation (\ref{eq:bhaml3.10}) vanishes in the mean
field approximation for the bubble system.

Let us now pay attention to the fluctuations with respect to the obtained
large number limits. Such fluctuations might appear for $\kappa =\gamma /\nu
\rightarrow 0$ in the large time scale $t\sim 1/\kappa $. We can get these
fluctuations without rescaling the time $t$ if we assume that $q_0=0$ and $%
\kappa =\gamma /\sqrt{\nu }$, so that we have to take into account also the $%
\kappa ^2$-terms in (\ref{eq:bhaml4.1}).

It follows from the Fock space representation (\ref{eq:bhaml4.6}) that the
quantum central limit 
\[
\lim_{\nu \rightarrow \infty }\frac{l}{\sqrt{\nu }}\,\hat{N}%
_{t}[L]=A_{t}[f_{0}^{\dagger }L]+A_{t}^{\dagger }[Lf_{0}] 
\]%
exists for any single-bubble operator $L$ with zero mean value $%
(f_{0},Lf_{0})=0$. We first apply this central limit theorem to the
right--hand side in (\ref{eq:bhaml2.2}): 
\begin{equation}
{\rm d}\psi (t)+K_{0}\psi (t){\rm d}t=\frac{{\rm i}}{\hbar }\,\gamma
(R\otimes {\rm d}\hat{u}_{t})\psi (t)\,.  \label{eq:bhaml4.8}
\end{equation}%
Here $K_{0}=K\otimes \hat{1}$, $K=\frac{{\rm i}}{\hbar }\,H+\frac{1}{2}%
\biggl(\frac{\gamma }{\hbar }\biggr)^{2}R\sigma ^{2}R$, 
\[
\hat{u}_{t}=A_{t}[f_{0}^{\dagger }Q]+A_{t}^{\dagger }[Qf_{0}]=2\Re
A_{t}^{\dagger }[Qf_{0}] 
\]%
is the Fock space representation of the Wiener process $u_{t}$ with the
dispersion $\sigma ^{2}=\hbar ^{2}(\partial f_{0},\partial f_{0})$, defined
by the quantum stochastic multiplication formula \cite{bib:DM3,bib:12} 
\[
{\rm d}\hat{u}{\rm d}\hat{u}={\rm d}A_{t}[f_{0}^{\dagger }Q]{\rm d}%
A_{t}^{\dagger }[Qf_{0}]=f_{0}^{\dagger }QQf_{0}{\rm d}t\,. 
\]%
The central limit equation (\ref{eq:bhaml4.8}) for the unitary evolution of
the coupled system turns out to be a stochastic Schr\"{o}dinger-Ito equation
of diffusive type driven by the Wiener process $u_{t}$. The same conclusion
obviously holds for the $M$-particle system driven by $M$ independent Wiener
processes $\hat{u}_{t}(k)$ identical to $\hat{u}_{t}$, $k=1,\ldots ,M$: 
\[
{\rm d}\psi ^{M}(t)+K_{0}^{M}\psi (t){\rm d}t=\frac{{\rm i}}{\hbar }\,\gamma %
\biggl(\sum_{k=1}^{M}R(k)\otimes {\rm d}\hat{u}_{t}(k)\biggr)\psi ^{M}(t)\,, 
\]%
where $K^{M}=\frac{{\rm i}}{\hbar }\,H^{M}+\frac{1}{2}\biggl(\frac{\gamma }{%
\hbar }\,\biggr)^{2}\sum_{k=1}^{M}R(k)\sigma ^{2}R(k)$, $K_{0}^{M}=K^{M}%
\otimes \hat{1}$.

The application of the central limit theorem to the r.h.s. of the reduction
equation (\ref{eq:bhaml2.10}) yields an essentially different type of the
stochastic evolution, 
\begin{equation}
{\rm d}\chi (t)+K\chi (t){\rm d}t=\gamma R\chi (t){\rm d}\hat{v}_{t}\,,
\label{eq:bhaml4.9}
\end{equation}%
originally derived in \cite{bib:DM3} by quantum calculus method. Here $v_{t}$
is a complex Wiener vector-process, with the Fock-space representation 
\begin{equation}
\hat{v}_{t}=A_{t}[f_{0}^{\dagger }L^{\prime }]+A_{t}^{\dagger }[L^{\prime
}f_{0}]=\Re A_{t}^{\dagger }[(w_{0}+\bar{w}_{0})f_{0}]+{\rm i}\Im
A_{t}^{\dagger }[(w_{0}-\bar{w}_{0})f_{0}]
\end{equation}%
given by the complex osmotic velocity $w_{0}(\lambda )=-\partial \ln
f_{0}(\lambda )$ of a single bubble, and the operator $K$ is essentially the
same as in (\ref{eq:bhaml4.8}). The linear stochastic equation (\ref%
{eq:bhaml4.9}) has a unique solution $\chi (t,v)=T(t,v)\eta $ for a given $%
\chi (0,v)=\eta \in {\cal H}$ which is not normalized $\Vert \chi (t,v)\Vert
\neq 0$ for every Wiener trajectory $t\mapsto v_{t}$ but is normalized in
the mean square sense $\int \Vert \chi (t,v)\Vert ^{2}{\rm P}_{0}({\rm d}%
v)=1 $ with respect to the Gaussian probability measure ${\rm P}_{0}$ of $%
v=\{v_{t}|t>0\}$. The measure ${\rm P}_{0}$ is defined by the zero mean
values of $v_{t}$ and by the table 
\begin{eqnarray*}
{\rm d}\hat{v}{\rm d}\hat{v} &=&{\rm d}A_{t}[f_{0}^{\dagger }L^{\prime }]%
{\rm d}A_{t}^{\dagger }[L^{\prime }f_{0}]=f_{0}^{\dagger }L^{\prime
}L^{\prime }f_{0}{\rm d}t \\
{\rm d}\hat{v}^{\ast }{\rm d}\hat{v} &=&{\rm d}A_{t}[f_{0}^{\dagger
}L^{\prime \dagger }]{\rm d}A_{t}^{\dagger }[L^{\prime
}f_{0}]=f_{0}^{\dagger }L^{\prime \dagger }L^{\prime }f_{0}{\rm d}t
\end{eqnarray*}%
of commuting multiplication 
\[
{\rm d}\hat{v}{\rm d}\hat{v}^{\ast }={\rm d}\hat{v}^{\ast }{\rm d}\hat{v}\,. 
\]%
But the reduction noise $\hat{v}_{t}$ obtained in the Fock space
representation does not commute with the real Wiener process $\hat{u}_{t}=%
\hat{u}_{t}^{\ast }$ in (\ref{eq:bhaml4.8}), 
\begin{eqnarray*}
{\rm d}\hat{v}{\rm d}\hat{u} &=&{\rm d}A_{t}[f_{0}^{\dagger }L^{\prime }]%
{\rm d}A_{t}^{\dagger }[Qf_{0}]=f_{0}^{\dagger }L^{\prime }Qf_{0}{\rm d}t \\
{\rm d}\hat{u}{\rm d}\hat{v} &=&{\rm d}A_{t}[f_{0}^{\dagger }Q]{\rm d}%
A_{t}^{\dagger }[L^{\prime }f_{0}]=f_{0}^{\dagger }QL^{\prime }f_{0}{\rm d}t
\end{eqnarray*}%
if $g(\lambda )\equiv \partial ^{2}\ln f_{0}(\lambda )\neq 0$, since 
\[
\lbrack {\rm d}\hat{v},{\rm d}\hat{u}]=f_{0}^{\dagger }[L^{\prime },Q]f_{0}%
{\rm d}t=\frac{\hbar }{{\rm i}}\,(f_{0},gf_{0}){\rm d}t\,. 
\]

In the same way one can obtain the continuous reduction equation 
\begin{eqnarray}
\lefteqn{{\rm d}\varrho^M(t)+(K\varrho^M(t)+\varrho^M(t)K^\dagger){\rm d} t=}
\nonumber \\
&& \biggl(\frac\gamma\hbar\biggr)^2\!\sum_{k=1}^M R(k)\sigma^2 \varrho^M(t)
R(k){\rm d} t+\gamma({\rm d} w_t R\varrho^M(t)+ \varrho^M(t) R{\rm d} w^*_t)
\label{eq:bhaml4.10}
\end{eqnarray}
where $R=\frac 1M\sum_{k=1}^MR(k)$ and the operator $K=K^M$ is the same as
in the equation for the unitary evolution of the $M$-particle system coupled
with the bubbles. The derived stochastic equation for the $M$-particle
density operator $\varrho^M(t,w)$ normalized in the mean is driven by the
complex Wiener process $w_t=v_t^M$ having the same multiplication table as
the process $\sqrt{M}v_t$ in (\ref{eq:bhaml4.9}). The diffusive type
equation (\ref{eq:bhaml4.10}), (\ref{eq:bhaml3.10}) also has the mixing
property. It was also derived in \cite{bib:ref33} by operational method.

\section{Conclusion}

\typeout{Conclusion} \label{sec:bham5}

The projection postulate for the reduction process follows as a result of
conditioning by the results of a nondemolition measurement in the von
Neumann Hamiltonian model \cite{bib:bhaml6} for the particle-meter
interaction. This model gives also all unsharp reductions for the
measurements with continuous spectrum.

The singular interaction, corresponding to the collision model of
scatterings can be treated in quantum mechanics also in terms of
Schr\"odinger equation in the generalized sense. The spontaneous
localization \cite{bib:bel7,bib:bel10} of the particle under the continual
observation is explained as the result of the nondemolition countings of the
scattered bubbles in a measurement apparatus like cloud chamber.

The spontaneous localization of a system of the identical particles (atoms)
in a bubble chamber (photodetector) is mixing due to the
indistinguishability of the particles via the measurements of the bubbles.
It is described in terms of the filtering equation for the density matrix of
the particles, driven by the Poisson process of the total scatterings of the
bubbles. This equation was derived in \cite{bib:ref33} within the
operational approach to the quantum continual measurements.

The macroscopic limit of the spontaneous localization equation is described
in terms of the Schr\"odinger one, corresponding to the mean field
approximation. The central limit of the spontaneous localization equation
proves the diffusive reduction equation, which appeared in \cite%
{bib:bhaml17,bib:bhaml18} for some cases of $R$ and $H$. Our theory shows
the origin of the Wiener processes generating the Ito diffusion equations
for quantum states \cite{bib:bhaml19, bib:bhaml20}, and also derives a
diffusive mixing equation for the reduced density matrix of the system of
particles under the continual observation.

The quantum jump model, based on Poisson distributed impulsive measurements,
is not new, see for example \cite{bib:bhaml21, bib:bhaml22}, and it has been
described on the statistical (a priori) level by the master equation which
also results from the quantum diffusion models. However the treatment of
these processes on the individual (a posteriori) level gains different types
of quantum dynamics which can be described only in terms of the stochastic
differential equations. The first such treatment of the counting
measurements and the a posteriori dynamics was suggested in \cite{bib:ref19,
bib:ref21} in terms of quantum stochastic calculus method \cite{bib:12}. The
present work puts this method on the solid mathematical basis by treatment
of $\delta $-function couplings with the classical Ito calculus applied to
the quantum jumps and spontaneous localizations.

{\bf Acknowledgements:} The first author (VPB) is greatfull for hospitality
of Marburg University during this work in the summer 1991 when he hold Guest
Professorship supported by Deutsche Forschungsgemeinschaft.

\end{document}